\documentclass[aps,prb,reprint,superscriptaddress]{revtex4-2}

\usepackage{graphicx}
\usepackage{amsmath}
\usepackage{amssymb}
\usepackage{physics}
\usepackage{bm}
\usepackage{xcolor}
\usepackage{csquotes}
\usepackage{hyperref}
\hypersetup{
  colorlinks=true,
  breaklinks=true,
  linkcolor=blue,
  urlcolor=blue,
  citecolor=blue,
}
\usepackage{changes}

\def\<#1>{\mathinner{\langle#1\rangle}}
\mathchardef\up="0222
\mathchardef\dn="0223

\makeatletter
\renewcommand*\env@matrix[1][\arraystretch]{%
  \edef\arraystretch{#1}%
  \hskip -\arraycolsep
  \let\@ifnextchar\new@ifnextchar
  \array{*\c@MaxMatrixCols c}}
\makeatother

\begin{document}
\preprint{arXiv}

\title{Coherent cellular dynamical mean-field theory: a real-space quantum embedding approach to disorder in strongly correlated electron systems}

\author{Patrick Tscheppe}
\email{ptscheppe@flatironinstitute.org}
\affiliation{Max-Planck-Institut für Festkörperforschung, Heisenbergstraße 1, 70569 Stuttgart, Germany}
\affiliation{Department of Physics, Columbia University, 538 West 120th Street, New York, New York 10027, USA}
\affiliation{Center for Computational Quantum Physics, Flatiron Institute, 162 5th Avenue, New York, New York 10010, USA}

\author{Marcel Klett}
\affiliation{Max-Planck-Institut für Festkörperforschung, Heisenbergstraße 1, 70569 Stuttgart, Germany}

\author{Henri Menke}
\affiliation{Max-Planck-Institut für Festkörperforschung, Heisenbergstraße 1, 70569 Stuttgart, Germany}

\author{Sabine Andergassen}
\affiliation{Institute of Information Systems Engineering, TU Wien, 1040 Vienna, Austria}
\affiliation{Institute of Solid State Physics, TU Wien, 1040 Vienna, Austria}

\author{Niklas Enderlein}
\affiliation{Department of Physics, Friedrich-Alexander-Universit\"at Erlangen/N\"urnberg, 91058 Erlangen, Germany}

\author{Philipp Hansmann}
\affiliation{Department of Physics, Friedrich-Alexander-Universit\"at Erlangen/N\"urnberg, 91058 Erlangen, Germany}

\author{Thomas Schäfer}
\email{t.schaefer@fkf.mpg.de}
\affiliation{Max-Planck-Institut für Festkörperforschung, Heisenbergstraße 1, 70569 Stuttgart, Germany}

\date{\today}

\begin{abstract}
We formulate a quantum embedding algorithm in real-space for the simultaneous theoretical treatment of nonlocal electronic correlations and disorder, the coherent cellular dynamical mean-field theory (C-CDMFT). This algorithm  combines the molecular coherent potential approximation with the cellular dynamical mean-field theory. After a pedagogical review of quantum embedding theories for disordered and interacting electron systems, and a detailed discussion of its work flow, we present first results from C-CDMFT for the half-filled two-dimensional Anderson-Hubbard model on a square lattice: (i) the analysis of its Mott metal-insulator transition as a function of disorder strength, and (ii) the impact of different types of disorder on its magnetic phase diagram. For the latter, by means of a ``disorder diagnostics'', we are able to precisely identify the contributions of different disorder configurations to the system's magnetic response.
\end{abstract}

\maketitle

\tableofcontents

\section{Introduction}
\label{sec:intro}

Despite ongoing advances in experimental instrumentation and synthesis methods, disorder~\cite{Lee1985,Alloul2009,Vojta2019} remains an intrinsic aspect in many quantum materials. Examples where disorder effects can play a significant role in the description and understanding of quantum matter are therefore vast, with effects leading to sometimes drastic, sometimes counterintuitive consequences: the (deliberate) introduction of disorder can heavily affect the superconducting transition temperatures $T_c$ in monolayers of 2H-TaS$_2$~\cite{Peng2018} and cuprates~\cite{Leroux2019,Alloul2024}, it can shift the metal-insulator boundary in organic charge-transfer salts~\cite{Gati2018,Riedl2022}, it may be responsible for the enhancement of the magnetic response in the parent compound of infinite-layer nickelates~\cite{Ortiz2022,Klett2022}, and be the driving force for nontrivial topological properties~\cite{Li2009,Silva2024}. These examples demonstrate that, depending on the specific case, disorder effects can be leveraged to enhance desired material properties~\cite{Mazza2024}, rather than being merely detrimental.

Recently, ultracold atomic quantum simulators have emerged as another highly controllable platform for the study of disorder effects and interactions \cite{Roati2008, Billy2008, White2009, Kondov2015}, including direct investigations of many-body localization \cite{Schreiber2015, Rispoli2019}.

Many of these fascinating examples, however, also highlight the challenge of fully describing disorder effects in these systems: disorder and (often nonlocal) correlations have to be treated on an equal footing for a comprehensive understanding of their physics. These two ingredients can be represented with the minimal model
\begin{equation}\label{eq:H_AndersonHubbard}
     H = \sum_{ij\sigma} t_{ij} c^\dagger_{i\sigma} c_{j\sigma}  + U\sum_{i} n_{i\up} n_{i\dn} + \sum_{i\sigma} \qty(\varepsilon_i - \mu) n_{i\sigma},
\end{equation}
where $t_{ij}$ represent the tunneling (hopping) amplitudes from lattice site $i$ to site $j$, $c^{(\dagger)}_{i\sigma}$ fermionic annihilation (creation) operators, respectively, for site $i$ and spin $\sigma$, $n_{i\sigma}$ the number operator, $U$ the local Coulomb repulsion strength, $\mu$ the chemical potential, and $\varepsilon_i$ an onsite (local) potential. If the local potential $\varepsilon_i$ and/or the transfer integrals $t_{ij}$ are treated as random variables, Eq.~(\ref{eq:H_AndersonHubbard}) is the well-known Anderson-Hubbard Hamiltonian~\cite{Anderson1958, Hubbard1963}.

\begin{figure}[t!]
    \centering
    \includegraphics{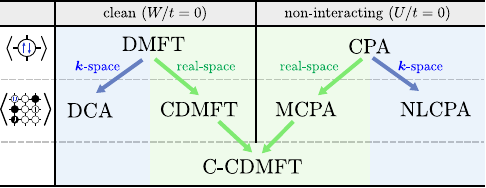}
    \caption{Overview of quantum embedding methods for clean interacting, and disordered noninteracting systems, respectively. C-CDMFT combines the real-space methods CDMFT and MCPA, see main text.}
    \label{fig:disorder_methods}
\end{figure}

Already at the level of the noninteracting disordered model ($U=0$, Anderson model), the possibility of Anderson localization~\cite{Anderson1958, Abou-Chacra1973} opens the path to nontrivial metal-insulator transitions.
Crucially, the average density of states does not become critical across this transition --- no single-particle gap is formed --- and in principle it becomes necessary to study the full statistical distribution of these functions. Including interactions in such a framework, known as statistical dynamical mean-field theory (statDMFT)~\cite{Dobrosavljevic1997, Dobrosavljevic1998}, quickly becomes prohibitively expensive, except for the simplest approximations.
A significant simplification was achieved with the development of the typical medium theory (TMT)~\cite{Dobrosavljevic2003, Dobrosavljevic2010, Miranda2012, Byczuk2005, Braganca2015} and extensions~\cite{Ekuma2014a, Ekuma2014b, Tam2021, Terletska2021}, in which the central quantity is not the average density of states, but rather the most likely (i.e., typical) one.
Referring to the extensive literature on this topic \cite{Lee1985, Kramer1993, Evers2008}, in this work we will not be addressing strong localization effects. Rather, we are interested in computing disorder averages of local observables --- these are the quantities that are measured in experiments.
A variety of reliable computational approaches have been developed for this problem. Most notably the coherent potential approximation (CPA,~\cite{Soven1967, VelickyKirkpatrickEhrenreich1968, YonezawaMorigaki1973}) and its cluster extensions: the molecular CPA (MCPA,~\cite{Tsukada1969, Ducastelle_1974}) in real-space and the dynamical cluster approximation (DCA\cite{Jarrell2001, Rowlands2006, Rowlands_2009}, but here referred to as NLCPA, following~\cite{Rowlands2006}, to avoid confusion with a similar method developed for periodic systems) in momentum space. A different formulation of this approach has recently been investigated~\cite{Moradian2019,Moradian2020}.

An additional complication arises from the possibility of \textit{off-diagonal} disorder, i.e., when impurities affect both the onsite energies and hopping amplitudes. This can be dealt with by a  Blackman-Esterling-Berk (BEB,~\cite{BEB1971, Esterling1975}) reformulation of the above methods.

On the other hand, in the limit of clean interacting Hubbard models~\cite{Hubbard1963,Qin2022,Arovas2022}, a powerful and well-established framework is provided by the dynamical mean-field theory (DMFT,~\cite{MetznerVollhardt1989, Georges1992, Georges1996}) and its respective cluster extensions: cellular DMFT (CDMFT,~\cite{Kotliar2001, Maier2005}) in real and the dynamical cluster approximation (DCA,~\cite{Hettler1998, Hettler2000,Maier2005}) in momentum space. Fig.~\ref{fig:disorder_methods} shows a conceptual overview of these methods, their realm of applicability, and their relations to each other.

The simultaneous treatment of electronic correlations and disorder poses a significant challenge to contemporary condensed matter theory (see \cite{Roemer2001, Miranda2005} and references therein). Strategies for the incorporation of onsite and off-diagonal disorder that build upon quantum embedding theories include DMFT+CPA~\cite{Weh2021}, auxiliary coherent potential approximation DMFT (ACPA-DMFT)~\cite{Zhang2025}, and the typical medium dynamical cluster approximation (TMDCA)~\cite{Ekuma2014a, Ekuma2014b, Terletska2014,Terletska2018,Zhang2016,Tam2021}.

In this manuscript we complement these approaches by combining the MCPA in its BEB formulation for treating disorder and CDMFT to handle the interactions. This quantum cluster theory, which we coin coherent CDMFT (C-CDMFT), is formulated entirely in real-space. It treats fluctuations stemming from the mutual electron-electron interactions and impurity scattering up to the size $N_c$ of the cluster exactly and on an equal footing. Also, it is a controlled technique, i.e., C-CDMFT approaches the exact solution of the model as $N_c \rightarrow \infty$. As the method is formulated in real-space, (i) there are indications both from clean system calculations~\cite{Klett2020} and in the noninteracting disordered limit~\cite{Rowlands2006} that it has favorable convergence properties with $N_c$, (ii) we can straightforwardly apply translation-symmetry breaking external fields, and (iii) we are able to identify the contributions of different disorder configurations to the system's response to such applied fields (``disorder diagnostics''). After a pedagogical review of quantum embedding theories for disorder, we show first applications of C-CDMFT to the dirty Mott transition and to the antiferromagnetism of the half-filled Anderson-Hubbard model on a two-dimensional square lattice.

The manuscript is organized as follows: In Sec.~\ref{sec:non_interacting} we review embedding methods for the treatment of noninteracting disordered systems: the CPA (Sec.~\ref{sec:cpa}), the BEB for offdiagonal disorder (Sec.~\ref{sec:beb}), as well as MCPA (Sec.~\ref{sec:mcpa}). In Sec.~\ref{sec:cdmft} we contrast these methods to DMFT and CDMFT approaches for clean systems. Our new results are presented in Secs.~\ref{sec:ccdmft}-\ref{sec:afm}. In Sec.~\ref{sec:ccdmft} we combine the BEB-MCPA and CDMFT into C-CDMFT and discuss in detail its algorithmic flow. We then present first results of the new method, first for the dirty Mott transition in Sec.~\ref{sec:mott}, and then for the antiferromagnetic phase of the Anderson-Hubbard model in Sec.~\ref{sec:afm}, before we conclude in Sec.~\ref{sec:conclusions}.

\begin{figure*}
    \centering
    \includegraphics[width=1\textwidth]{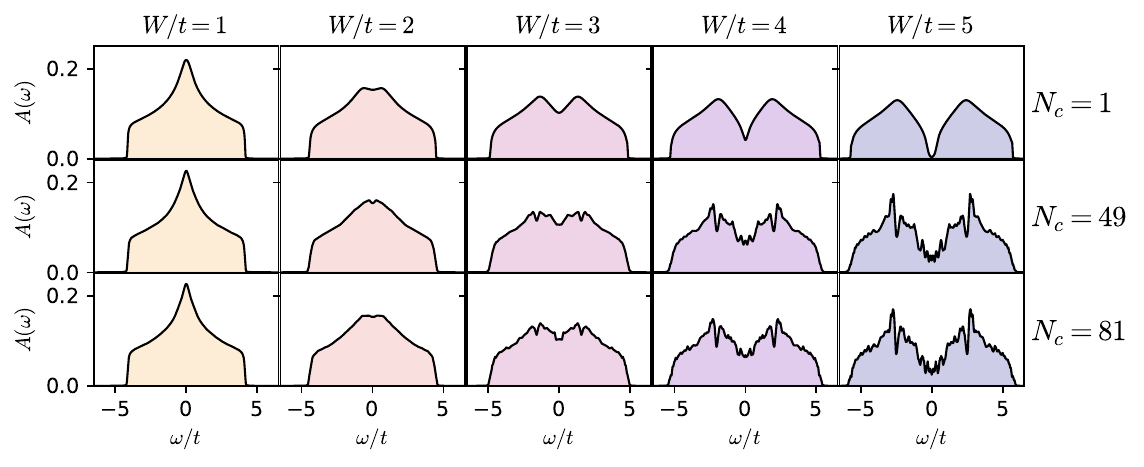}
    \caption{Local spectral functions $A(\omega)$ on the simple two-dimensional square lattice with impurity concentration $c=0.5$, for different disorder strengths $W/t$. The top panel shows CPA results ($N_c = 1$), and converged MCPA data for cluster sizes $N_c = 49$ and $81$ is also shown for comparison. We perform the configuration averaging using a stochastic sampling procedure (Appendix \ref{app:Sampling}).}
    \label{fig:2d_MCPA_spectra}
\end{figure*}

\section{Review: Noninteracting Electrons in Disordered Lattices}
\label{sec:non_interacting}
Consider first the noninteracting limit of Hamiltonian Eq.~(\ref{eq:H_AndersonHubbard}), the Anderson model~\cite{Anderson1958}
 \begin{equation}\label{eq:H_AndersonModel}
     H = \sum_{ij\sigma} t_{ij} c^\dagger_{i\sigma}c_{j\sigma} + \sum_{i\sigma} \qty(\varepsilon_i - \mu) n_{i\sigma}.
 \end{equation}
 Both the local potential $\varepsilon_i$ and transfer integrals $t_{ij}$ are treated as random variables, but we will restrict their probability distribution to that of a \emph{binary mixture} model: there are two inequivalent components, arranged randomly on the lattice with probability $c_\mathrm{A} \equiv 1 -  c$ (\enquote{A sites}) and $c_\mathrm{B} \equiv c$ (\enquote{B sites}); their onsite potentials are
 \begin{equation}\label{eq:random_potentials}
     \varepsilon_i = \begin{cases}-W/2 \qq{on A sites} \\
     +W/2 \qq{on B sites,}\end{cases}
\end{equation}
where the quantity $W > 0$ is a measure of the disorder potential strength.
Furthermore, the hopping amplitudes between sites may depend on the chemical composition of those sites. For nearest neighbor hoppings, a simple parametrization is $t_{ij} \tau^{\mathrm{IJ}}$, where $t_{ij}$ is the usual translation invariant part with value $-t$ on nearest neighbor bonds and
\begin{equation}\label{eq:random_hoppings}
     \tau^\mathrm{IJ} = \begin{cases}
         \tau^\mathrm{AA} \qq{for $i,j$ both A sites} \\
         \tau^\mathrm{AB} \qq{if $i, j \in \{ (\mathrm{A, B}), (\mathrm{B, A}) \}$} \\
         \tau^\mathrm{BB} \qq{for $i,j$ both $B$ sites.}
     \end{cases}
\end{equation}
This model is particularly fitting for a description of chemically doped compounds, where, say, $c_\mathrm{B}$ would correspond to the concentration of dopants introduced into the host material.
Situations in which fluctuations in either $\varepsilon_i$ or $\tau^{\mathrm{IJ}}$ only are considered are referred to as \emph{diagonal} and \emph{off-diagonal} disorder, respectively.

In the following sections the chemical potential will be absorbed into a redefinition of the energy levels $\varepsilon_i$ in order to simplify the expressions.

\subsection{Diagonal Disorder: Coherent Potential Approximation (CPA)}
\label{sec:cpa}
A natural framework for discussing various real-space approaches to the problem of disorder and interactions is the locator expansion~\cite{Anderson1958, Ziman1969, ElliottReview1974, Maier2005}, which takes as its basic building block the Green function of an isolated site $g_i = (\omega - \varepsilon_i)^{-1}$ and treats hopping between sites as a perturbation.
Accordingly, the full Green function can be written as
\begin{equation}\label{eq:locator_expansion_G}
    G_{ij} = g_i\,\delta_{ij} + g_i \, t_{ij} \, g_j + \sum_k g_i \, t_{ik} \, g_k \, t_{kj}  \, g_j + \dots,
\end{equation}
where each $g_i$ is a random variable, determined by the value of the energy $\varepsilon_i$ at site $i$.
We are interested in finding $\expval{G_{ij}}$, i.e., the Green function averaged over all microscopic realizations of disorder. This is not a straightforward task, as the number of such configurations grows exponentially with the system size. Instead, there is a family of approximate techniques of increasing accuracy, which we review in the following. We start with the CPA~\cite{Soven1967, VelickyKirkpatrickEhrenreich1968, YonezawaMorigaki1973}, which provides a good single-particle description at little computational cost, while already capturing nontrivial physics, such as the band splitting transition. As in DMFT, to be introduced later on, the CPA becomes exact in the limit of large lattice coordination~\cite{VlamingVollhardt1992}: %---
it includes all scattering processes caused by the \emph{local} potential fluctuations at a given site and treats the rest of the lattice in a self-consistent mean-field fashion (see~\cite{ElliottReview1974, YonezawaMorigaki1973} for a detailed discussion).

To be more precise, focus on the site-diagonal component $G_{ii}$ in Eq.~\eqref{eq:locator_expansion_G} and isolate all occurrences of the locator $g_i$ for that particular site:
\begin{equation}\label{eq:local_G_interactor_form}
    G_{ii} =  g_i  + g_i \Delta_{i} g_i + g_i \Delta_{i} g_i \Delta_{i} g_i + \dots = \frac{1}{g_i^{-1} - \Delta_{i}},
\end{equation}
where the function $\Delta_{i}(\omega)$ consists of all hopping processes starting and ending at site $i$ without returning to it at any intermediate step:
\begin{equation}\label{eq:CPA_interactor}
    \Delta_{i} = \sum_{j\neq i} t_{ij} g_j t_{ji} + \sum_{jk\neq i} t_{ij} g_j t_{jk} g_k t_{ki} + \dots.
\end{equation}
Having isolated the local contribution to $G_{ii}$, the CPA proceeds by replacing $\Delta_i$ with its average value $\Delta_i \approx \Delta^{\mathrm{CPA}}$. Averaging the Green function $\expval{G_{ii}}$ now involves only a single site:
\begin{subequations}

\begin{align}
    \expval{G_{ii}(\omega)} &= \expval{\frac{1}{g^{-1}_i(\omega) - \Delta^{\mathrm{CPA}}(\omega)}} = \label{eq:CPA_average_G}\\ &= \frac{c_\mathrm{A}}{\omega - \varepsilon_\mathrm{A} - \Delta^{\mathrm{CPA}}(\omega)} +\frac{c_\mathrm{B}}{\omega - \varepsilon_\mathrm{B} - \Delta^{\mathrm{CPA}}(\omega)}. \nonumber
\end{align}

To fix the value of $\Delta^{\mathrm{CPA}}$, the lattice environment of site $i$ is replaced by an effective, translation-invariant medium, parametrized by a locator $\gamma$, and chosen in such a way as
to self-consistently reproduce the local Green function $\expval{G_{ii}}$:
\begin{align}
    \expval{G_{ii}(\omega)} &= \frac{1}{\gamma^{-1}(\omega) - \Delta^{\mathrm{CPA}}(\omega)} \\
    &\overset{!}{=} \frac{1}{N}\sum_{\bm{k}} \frac{1}{\gamma^{-1}(\omega) - \varepsilon_{\bm{k}}}.
\end{align}
Introducing a (local) self-energy $\Sigma$ via $\gamma = (\omega - \Sigma(\omega))^{-1}$ brings this into the form
\begin{align}
    \expval{G_{ii}(\omega)} &= \frac{1}{\omega - \Sigma(\omega) - \Delta^{\mathrm{CPA}}(\omega)} \\
    &\overset{!}{=} \frac{1}{N}\sum_{\bm{k}} \frac{1}{\omega - \Sigma(\omega) - \varepsilon_{\bm{k}}}\label{eq:CPA_self-consistency},
\end{align}
\end{subequations}
which is structurally identical to the DMFT (see Sec.~\ref{sec:cdmft}).
Equations~\eqref{eq:CPA_average_G}-\eqref{eq:CPA_self-consistency} need to be solved in tandem. Their solution determines both the average local Green function $\expval{G_{ii}}$ as well as the auxiliary quantities $\Sigma$ and $\Delta^{\mathrm{CPA}}$.

This single-site CPA solution reproduces many physically desirable features: it becomes exact in the dilute limit of either species, $c_\mathrm{A} \ll 1$ or $c_\mathrm{B} \ll 1$, as well as the weakly disordered limit $W/t \ll 1$, and the atomic limit $t \rightarrow 0$. As $W$ exceeds the noninteracting bandwidth, only sites with the lower potential will be (doubly) occupied, leaving the remaining sites approximately empty, and turning the system into a band insulator. This splitting of the density of states into impurity subbands is captured by the CPA~\cite{YonezawaMorigaki1973}, as illustrated in Fig.~\ref{fig:2d_MCPA_spectra} (top row), which shows typical CPA spectral functions $A(\omega) = - \frac{1}{\pi} \Im \expval{G_{ii}(\omega + i0^+)}$ for various values of $W/t$ and a symmetric impurity concentration $c_\mathrm{A} = c_\mathrm{B} = 0.5$.
\begin{figure*}
    \centering
    \includegraphics[width=0.9\textwidth]{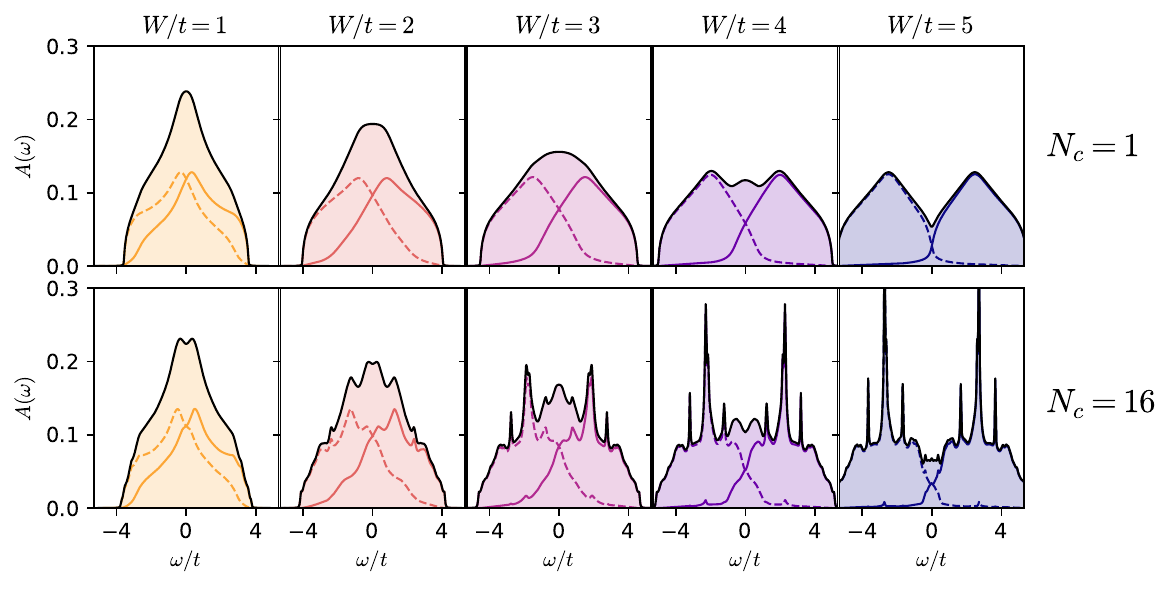}
    \caption{BEB ($N_c = 1$, top row) and BEB-MCPA ($N_c = 16$, bottom row) spectral functions $A(\omega)$ (black) for the simple square lattice with both diagonal and off-diagonal disorder. The component-resolved densities of states $A^\mathrm{A}(\omega)$ (dashed) and $A^\mathrm{B}(\omega)$ (solid) are shown in color. The parameters are $c=0.5$, $\tau^\mathrm{AA} = \tau^\mathrm{BB} = 1$, $\tau^\mathrm{AB} = 1/2$ and various values of $W/t$. % are considered.
    }
    \label{fig:2dMCPA-BEB_spectra}
\end{figure*}
These figures also reveal a fundamental flaw of the single-site approach: due to the relatively simple analytic structure of $\expval{G_{ii}}$, the spectra are %seen to be
very smooth. This is in contrast to exact calculations of the density of states in model systems, which often feature a complicated peak structure as well as exponentially small tails at the band edges~\cite{Dean1972, Lifshitz1965}.
Furthermore, no direct information about localization can be obtained within the CPA itself. In particular, Anderson's localization transition is absent~\cite{Haydock1974, Jarrell2001, Miranda2012}.

\subsection{Off-diagonal Disorder: Blackman-Esterling-Berk (BEB) Formalism}
\label{sec:beb}

Even though the CPA is a local approximation formulated for local impurity potentials $\varepsilon_i$, more general impurity potentials can be straightforwardly included. For a more realistic description of disorder in real materials an important effect is that the hopping amplitudes between sites may also depend on the types of atoms occupying these sites ((A, A), (B, B) or (A, B)). Since hopping to and away from a site $i$ is encoded in the quantity $\Delta_i$, it is clear that the CPA replacement $\Delta_i \rightarrow \Delta^{\mathrm{CPA}}$ is too restrictive. Instead, a different effective medium is required for each atomic species, so that a distinction between A and B sites can be maintained throughout the averaging procedure.

As demonstrated by Blackman, Esterling and Berk~\cite{BEB1971, Esterling1975}, such a procedure can be carried out by reintroducing a component label $\mathrm{I} \in \{ \mathrm{A}, \mathrm{B} \}$ on each site, generalizing the scalar CPA quantities $g_i$, $\Delta^{\mathrm{CPA}}$, and $\Sigma$ to matrices (in the following denoted by a hat). For this purpose it is convenient to introduce projectors onto the local atomic species $\dyad{A_i}{A_i}$ and $\dyad{B_i}{B_i}$, and to expand quantities in this basis, e.g.,
\begin{subequations}
\begin{align}
    \Hat{t}_{ij} &= \sum_{IJ} \dyad{I_i}t_{ij}{\dyad{J_j}} \equiv \sum_{IJ} \ket{I_i}t^{\mathrm{IJ}}_{ij}{\bra{J_j}} \label{eq:t_BEB_mat}\\
    \Hat{g}_{i} &\equiv \sum_{I} \ket{I_i}g^{\mathrm{I}}_{i}{\bra{I_i}}, \label{eq:g_BEB_mat}\\
    \Hat{G}_{ij} &\equiv \sum_{IJ} \ket{I_i}G^{\mathrm{IJ}}_{ij}{\bra{J_j}}, \label{eq:G_BEB_mat}
\end{align}
\end{subequations}
where the matrix elements $t_{ij}^{\mathrm{IJ}} \equiv t_{ij} \tau^{\mathrm{IJ}}$ are the same as in Eq.~\eqref{eq:random_hoppings}, and
\begin{equation}\label{eq:BEB_g_i_explicit}
    g_i^{\mathrm{I}} \equiv \begin{cases} (\omega - \varepsilon_\mathrm{I})^{-1} &\qq{if $i$ is an I site} \\
    0 &\qq{otherwise.}
    \end{cases}
\end{equation}

The locator expansion of $\Hat{G}_{ij}$ then becomes identical in form to Eq.~\eqref{eq:locator_expansion_G}, i.e.,
\begin{align}
    \Hat{G}_{ij} =  \Hat{g}_i\,\delta_{ij} + \Hat{g}_i \, \Hat{t}_{ij} \, \Hat{g}_j + \sum_k \Hat{g}_i \, \Hat{t}_{ik} \, \Hat{g}_k \, \Hat{t}_{kj}  \, \Hat{g}_j + \dots,
\end{align}
with implied matrix multiplication. % is implied.
The merit of this rewriting is that the only random quantity occurring in the expansion is the generalized locator $\hat{g}_i$. It is the locator in Eq.~\eqref{eq:BEB_g_i_explicit} that picks out the correct component of the hopping based on the values of the local potential.

The BEB-CPA equations can now be written in complete analogy to Eqs.~\eqref{eq:CPA_average_G}-\eqref{eq:CPA_self-consistency}~\cite{BEB1971}\footnote{Note that the quantities $\Hat{g}_i$ are not invertible. However, averages over disorder are.}:
\begin{subequations}
\begin{align}
    \expval{\Hat{G}_{ii}(\omega)} &= \expval{\qty[\Hat{1} - \Hat{g}_i(\omega) \Hat{\Delta}^\mathrm{CPA}(\omega)]^{-1} \Hat{g}_i(\omega)} \label{eq:BEB1} \\
    &= \qty[\omega \Hat{1} - \Hat{\Sigma}(\omega) - \Hat{\Delta}^\mathrm{CPA}(\omega)]^{-1} \label{eq:BEB2}\\
    &= \frac{1}{N} \sum_{\bm{k}} \qty[\omega \Hat{1} - \Hat{\Sigma}(\omega) - \Hat{\tau} \varepsilon_{\bm{k}}]^{-1}. \label{eq:BEB3}
\end{align}
\end{subequations}
The single-site average in Eq.~\eqref{eq:BEB1} can be performed explicitly and yields
\begin{equation}\label{eq:BEB_explicit}
    \expval{\Hat{G}_{ii}(\omega)} = \begin{pmatrix}
        \dfrac{c_\mathrm{A}}{\omega - \varepsilon_\mathrm{A} - \Delta^\mathrm{AA}(\omega)} & 0 \\
        0 & \dfrac{c_\mathrm{B}}{\omega - \varepsilon_\mathrm{B} - \Delta^\mathrm{BB}(\omega)}
    \end{pmatrix}.
\end{equation}
Physically, the diagonal component $\langle G^{\mathrm{II}}_{ii} \rangle$  corresponds to the local Green function of an atom of type I embedded in an effective medium obtained by averaging over the rest of the lattice. Then the full average is simply the sum
\begin{equation}\label{eq:BEB_full_avg}
    \expval{G_{ii}(\omega)} = \expval{G^\mathrm{AA}_{ii}(\omega)} + \expval{G^\mathrm{BB}_{ii}(\omega)},
\end{equation}
which, using Eq.~\eqref{eq:BEB_explicit}, reduces to Eq.~\eqref{eq:CPA_average_G} in the CPA limit $\Delta^{\mathrm{AA}} = \Delta^{\mathrm{BB}}$, i.e., when the coupling to the medium is identical for both species.

The spectral functions obtained using this formalism are displayed in the top row of Fig.~\ref{fig:2dMCPA-BEB_spectra}, which shows both the full average and the component-resolved densities of states. The local potential and concentration are chosen as in Fig.~\ref{fig:2d_MCPA_spectra}, but hopping between A and B sites is suppressed by setting $\tau^{\mathrm{AB}} = 1/2$. This leads to a weaker hybridization between the impurity bands, partially filling in the gap at the largest potential strength $W/t = 5$.

\subsection{Nonlocal Correlations: Molecular Coherent Potential Approximation (MCPA)}
\label{sec:mcpa}

Having discussed formalism that include both onsite and hopping disorder, we now turn our attention to a quantitative improvement of the theory. The crucial flaw of all single-site approximations is their neglect of coherent scattering between \emph{different} lattice sites \footnote{This is a necessary ingredient for both Anderson localization~\cite{Anderson1958, Abou-Chacra1973} and the formation of Lifshitz tails~\cite{Lifshitz1965}}.
Various routes can be taken for going beyond the CPA, most notably quantum cluster methods~\cite{Maier2005} and diagrammatic extensions~\cite{Terletska_2013}. The former comes in two main flavors: a coarse-graining procedure in momentum-space, the NLCPA~\cite{Jarrell2001, Rowlands2006, Rowlands_2009} (commonly DCA, see the remark in Sec.~\ref{sec:intro}), and a direct real-space approach known as the MCPA~\cite{Tsukada1969, Ducastelle_1974}. In this work we are primarily concerned with studying local properties, such as the average density of states and staggered susceptibility, for which the MCPA scheme is generally better suited~\cite{Rowlands_2009}. We briefly review the construction of this method below, closely following Ref.~\cite{Maier2005}.

\begin{figure}
    \centering
    \includegraphics[width=0.3\textwidth]{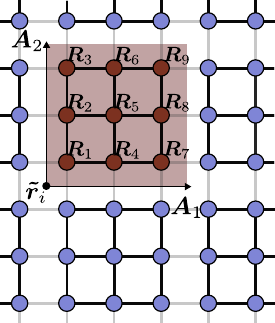}
    \caption{Example of a square $N_c = 3 \times 3 = 9$ lattice tiling. The origin of each cluster is labeled by a superlattice vector $\Tilde{\bm{r}_i} = n_i\, \bm{A}_1 + m_i \, \bm{A}_2$ with $n_i, m_i \in \mathbb{Z}$; sites at positions $\bm{r}_i = \Tilde{\bm{r}_i} + \bm{R}_i$ within each cell are numbered as shown.}
    \label{fig:CDMFT_cluster}
\end{figure}

For simplicity we first consider the case with diagonal disorder only. Begin by choosing a tiling of the lattice with identical clusters, each containing $N_c$ sites, and forming a superlattice with lattice vectors $\Tilde{\bm{r}}_i$. A lattice vector $\bm{r}_i$ in the original lattice can now be decomposed as $\bm{r}_i = \Tilde{\bm{r}}_i + \bm{R}_i$, where $\bm{R}_i$ labels lattice vectors on the cluster (see Fig.~\ref{fig:CDMFT_cluster}). With this notation at hand, the locator expansion can be rearranged in terms of the bare locators $\underline{g}_i$ of isolated clusters. These are $N_c \times N_c$ matrices (such matrices will be denoted by underlines to distinguish them from BEB quantities) of the form
\begin{equation}
    \underline{g}_i(\omega) = \qty[\omega \, \underline{1} - \begin{pmatrix}
        \varepsilon_1 & & \\
        & \ddots & \\
        & & \varepsilon_{N_c}
    \end{pmatrix} - \underline{t}_\mathrm{loc} ]^{-1},
\end{equation}
where the diagonals of $\underline{g}_i^{-1}$ contain the random site energies $\varepsilon_i$ and all intracluster hoppings are collected in the quantity $\underline{t}_\mathrm{loc}$. The full Green function matrix now depends only on the superlattice vectors $\Tilde{\bm{r}}_i, \Tilde{\bm{r}}_j$, and has matrix elements $[\underline{G}(\Tilde{\bm{r}}_i, \Tilde{\bm{r}}_j)]_{\bm{R}_i, \bm{R}_j} = G(\bm{r}_i, \bm{r}_j)$. Because intracluster hopping is already accounted for in $\underline{g}_i$, the locator expansion is now written in terms of the intercluster hopping $\underline{\delta t}$:
\begin{align}\label{eq:cluster_locator_expansion}
    \underline{G}(\Tilde{\bm{r}}_i, \Tilde{\bm{r}}_j) &= \underline{g}_i \delta_{\Tilde{\bm{r}}_i, \Tilde{\bm{r}}_j} + \underline{g}_i\,\underline{\delta t}(\Tilde{\bm{r}}_i - \Tilde{\bm{r}}_j)\,\underline{g}_j + \\ &+ \sum_k \underline{g}_{i}\,\underline{\delta t}(\Tilde{\bm{r}}_i - \Tilde{\bm{r}}_k)\,\underline{g}_k\,\underline{\delta t}(\Tilde{\bm{r}}_k - \Tilde{\bm{r}}_j)\,\underline{g}_j + \dots. \nonumber
\end{align}
The procedure is now identical to Eq.~\eqref{eq:local_G_interactor_form}. Pick out the cluster at position $\Tilde{\bm{r}}_i$ and bring its full Green function into the form
\begin{equation}\label{eq:cluster_G_interactor_form}
     \underline{G}(\Tilde{\bm{r}}_i, \Tilde{\bm{r}}_i) = \qty[\underline{g}_i^{-1} - \underline{\Delta}_i]^{-1}.
\end{equation}
The single-site CPA is now easily generalized by neglecting fluctuations in the quantity $\underline{\Delta}_i$. Then,  averaging the cluster Green function $\expval{\underline{G}(\Tilde{\bm{r}}_i, \Tilde{\bm{r}}_i)}$ involves only the sites on the cluster:
\begin{subequations}
\begin{equation}\label{eq:MCPA1}
    \expval{\underline{G}(\Tilde{\bm{r}}_i, \Tilde{\bm{r}}_i; \omega)} = \expval{\qty[\underline{g}_i^{-1}(\omega) - \underline{\Delta}^{\mathrm{CPA}}(\omega)]^{-1}},
\end{equation}
which is a sum over $2^{N_c}$ configurations. Introducing an effective medium to determine the value of $\underline{\Delta}^{\mathrm{CPA}}$ results in the two additional self-consistency constraints
\begin{align}
    \expval{\underline{G}(\Tilde{\bm{r}}_i, \Tilde{\bm{r}}_i; \omega)} &= \qty[\omega \underline{1} - \underline{\Sigma}(\omega) - \underline{\Delta}^{\mathrm{CPA}}(\omega)]^{-1} \label{eq:MCPA2}\\
    &\overset{!}{=} \frac{N_c}{N} \sum_{\Tilde{\bm{k}}} \qty[\omega \underline{1} - \underline{\Sigma}(\omega) - \underline{t}(\Tilde{\bm{k}})]^{-1}, \label{eq:MCPA3}
\end{align}
\end{subequations}
where the momentum $\Tilde{\bm{k}}$ runs over the superlattice Brillouin zone and $\underline{t}(\Tilde{\bm{k}})$ denotes the associated superlattice dispersion.

The set of equations Eqs.~\eqref{eq:MCPA1}-\eqref{eq:MCPA3} fully defines the MCPA. By construction, it reduces to the CPA for cluster size $N_c = 1$, and larger clusters increasingly take into account nonlocal scattering events on the cluster scale.
While the exact solution is recovered only in the limit $N_c \rightarrow \infty$, convergence to this result is typically fast enough that accurate estimates for disorder-averaged observables can already be obtained for accessible values of $N_c$~\cite{Rowlands_2009}.
To illustrate this point, in Fig.~\ref{fig:2d_MCPA_spectra} we show  local spectral functions computed on square clusters with $N_c = 1$ (CPA), $7\times 7 = 49$, and $9\times 9 = 81$.
While the CPA result already resembles that of the clusters for weak disorder, significant corrections to the simple single-site approximation become visible close to the band splitting transition.
Moreover, increasing the cluster size from 49 to 81 sites does not substantially alter the structure of the spectrum anymore, suggesting that their features remain robust in the limit $N_c \rightarrow \infty$ (see also Appendix \ref{app:1dresults} for a similar analysis of the one dimensional model).

Offdiagonal disorder can now be treated as before by reintroducing randomness into the hopping amplitudes $t_{ij}$. This is accomplished by augmenting each element of the cluster quantities with an additional BEB component structure, symbolically  $\underline{g} \rightarrow \underline{\Hat{g}}$. For example, the intracluster hopping can now be expressed as a direct product of the homogeneous cluster hopping $\underline{t}_{\mathrm{loc}}$ and the component matrix $\hat{\tau}$:  $\underline{\Hat{t}}_{\mathrm{loc}} = \underline{t}_{\mathrm{loc}} \otimes \Hat{\tau}$.
Following the same steps as before, we arrive at the self-consistency equations
\begin{subequations}
\begin{equation}\label{eq:ODDMCPA1}
    \expval{\underline{\Hat{G}}(\Tilde{\bm{r}}_i, \Tilde{\bm{r}}_i; \omega)} = \expval{\qty[\Hat{\underline{1}} - \underline{\Hat{g}}_i(\omega) \underline{\Hat{\Delta}}^\mathrm{CPA}(\omega)]^{-1}\underline{\Hat{g}}_i(\omega)},
\end{equation}
and
\begin{align}
    \expval{\underline{\Hat{G}}(\Tilde{\bm{r}}_i, \Tilde{\bm{r}}_i; \omega)}
    &= \qty[(\omega \, \Hat{\underline{1}} - \Hat{\underline{\Delta}}^\mathrm{CPA}(\omega) - \Hat{\underline{\Sigma}}(\omega)]^{-1} \label{eq:ODDMCPA2} \\
    &\overset{!}{=} \frac{N_c}{N} \sum\limits_{\Tilde{\bm{k}}} \qty[\omega \,\Hat{\underline{1}} - \underline{t}(\Tilde{\bm{k}}) \otimes \Hat{\tau} - \Hat{\underline{\Sigma}}(\omega)]^{-1} \label{eq:ODDMCPA3}.
\end{align}
\end{subequations}
These equations determine the cluster components $G_{ij}^{\mathrm{IJ}}$ at fixed chemical occupation of sites $i$ and $j$. The physical average is obtained by performing a sum over component indices, yielding
\begin{equation}\label{eq:BEB_MCPA_components_sum}
    \expval{G_{ij}(\omega)} = \expval{G_{ij}^{\mathrm{AA}}(\omega)} + \expval{G_{ij}^{\mathrm{AB}}(\omega)} + \expval{G_{ij}^{\mathrm{BA}}(\omega)} + \expval{G_{ij}^{\mathrm{BB}}(\omega)}.
\end{equation}
We note that off-diagonal terms like $\expval{G_{ij}^{\mathrm{AB}}}$ do not exist locally, i.e., for $i=j$. In this case Eq.~\eqref{eq:BEB_MCPA_components_sum} reduces to Eq.~\eqref{eq:BEB_full_avg}.

We can now study the effects of nonlocal scattering also in the case of off-diagonal disorder. Fig.~\ref{fig:2dMCPA-BEB_spectra} shows the corresponding spectral functions for two representative cluster sizes $N_c = 1$ (BEB-CPA) and $4 \times 4 = 16$, where the hopping between different species is suppressed ($\tau^{\mathrm{AB}} = 1/2$). In contrast to the case with purely diagonal disorder, we find that a suppression of the hybridization between components produces a very sharp peak structure already well before the band splitting transition occurs, i.e., the effects of short-ranged nonlocal fluctuations are much stronger in the case of offdiagonal disorder.

This ends our review of methods for noninteracting disordered systems. As we showed, the effective medium construction on which the CPA is based is sufficiently flexible to allow for the inclusion of both
extended impurity potentials, as well as nonlocal scattering effects.
The choice of such an embedding is not unique. While our aim is to describe local disorder averages, similar ideas have also been applied to the study of Anderson localization, both in real-space~\cite{Tam2021} and in momentum space~\cite{Terletska2014, Terletska2018, Zhang2016}.

\section{Review: Strongly Correlated Electrons in Clean Systems}
\label{sec:cdmft}

\begin{figure*}
    \centering
    \includegraphics[width=1\textwidth]{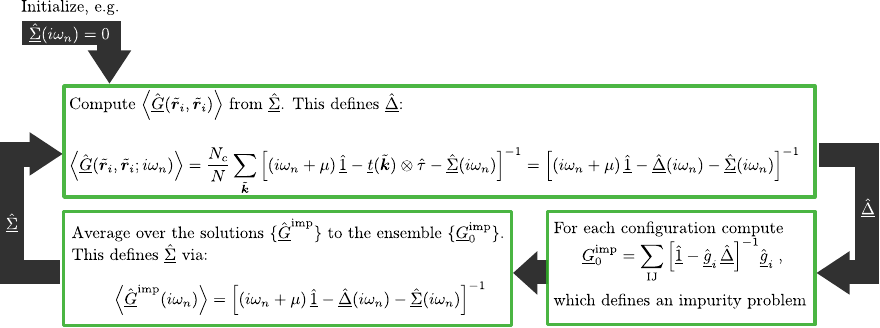}
    \caption{Self-consistency cycle of the C-CDMFT scheme. A BEB component structure in conjunction with the CDMFT mapping to an ensemble of Anderson impurity problems allows the combined treatment of electron interactions and disorder. At each step in the cycle, an ensemble of cluster impurity problems has to be solved, one for each unique configuration of the disorder potential.}
    \label{fig:SC_BEB-CDMFT}
\end{figure*}

The arguably most fundamental model for correlations in clean environments is the Hubbard model~\cite{Hubbard1963, Qin2022, Arovas2022}
\begin{equation}\label{eq:H_Hubbard}
     H = \sum_{ij\sigma} t_{ij} c^\dagger_{i\sigma}c_{j\sigma}  + U\sum_{i}n_{i\up} n_{i\dn} -\mu \sum_{i\sigma}  n_{i\sigma},
\end{equation}
which is the clean limit of Eq.~\eqref{eq:H_AndersonHubbard}.
We are interested in phenomena driven by strong interactions between the electrons, i.e., where $U$ is of the order of a material's bandwidth. In this regime a perturbative treatment of the interaction term is insufficient, but a local approximation of the CPA type has turned out to be very successful: this is the DMFT~\cite{MetznerVollhardt1989, Georges1992, Georges1996}. In complete analogy to the CPA, DMFT replaces the full lattice problem Eq.~\eqref{eq:H_Hubbard} by a single interacting site in an effective, noninteracting medium. Instead of performing a single-site average, this now requires solving for the Green function of a single-site quantum impurity model.
This impurity Green function can be written in the form
\begin{equation}
    G^{\mathrm{imp}}(i\omega_n) = \frac{1}{i\omega_n - \Delta^{\mathrm{imp}}(i\omega_n) - \Sigma^{\mathrm{imp}}(i\omega_n)},
\end{equation}
where the hybridization function $\Delta^{\mathrm{imp}}$ and self-energy $\Sigma^{\mathrm{imp}}$ play exactly the same roles as in Eq.~\eqref{eq:CPA_self-consistency}: $\Delta^{\mathrm{imp}}$ is determined by requiring that an effective medium characterized by the local self-energy $\Sigma^{\mathrm{imp}}$ reproduces the same local Green function as the quantum impurity model
\begin{equation}
    G^{\mathrm{imp}}(i\omega_n) \overset{!}{=} G_{ii}(i\omega_n) = \frac{1}{N} \sum_{\bm{k}} \frac{1}{i\omega_n - \varepsilon_{\bm{k}} - \Sigma^{\mathrm{imp}}(i\omega_n)}.
\end{equation}
While the local quantum fluctuations responsible for, e.g., the Mott transition~\cite{Hubbard1963, Mott1968, Imada_Review}, are successfully captured in this way, nonlocal correlations are completely ignored. %missing.
It is precisely those correlations that are responsible for some of the most interesting phenomena described by the repulsive Hubbard model, such as its pseudogap regime~\cite{Meixner2024,Simkovic2024} and superconductivity~\cite{Lichtenstein2000}. Even the DMFT picture of the Mott transition is substantially modified by the inclusion of such corrections~\cite{Park2008,Schaefer2015,Simkovic2020,Schaefer2021,Chatzieleftheriou2024}.

However, the theory can be refined in much the same way as in the disordered case: replacing the effective single-site impurity problem by a cluster of interacting sites gradually reintroduces the missing correlations. We will again focus on the direct real-space embedding of this cluster into the effective medium, mirroring the MCPA construction. The approximation obtained in this way is known as CDMFT~\cite{Kotliar2001, Maier2005} (see~\cite{Sakai2012, Klett2020} for detailed numerical studies on large clusters). We note that this choice of embedding is again not unique. The DCA~\cite{Hettler1998, Hettler2000,Maier2005} follows a complementary approach and is more naturally suited for the study of quantities in momentum-space.

In this work we solve the quantum impurity problem using the numerically exact continuous time quantum Monte Carlo algorithm in its interaction expansion form (CT-INT,~\cite{Rubtsov2004, Gull2011}), as implemented in the TRIQS software library~\cite{TRIQS}.

\section{Strongly Correlated Electrons in Disordered Lattices: the Coherent Cellular DMFT (C-CDMFT)}
\label{sec:ccdmft}

The following sections present the main novel contributions of this paper.
Having discussed reliable and systematic frameworks for the treatment of both disorder and electronic correlations separately, we are now in a position to combine the two and address the full Anderson-Hubbard Hamiltonian Eq.~\eqref{eq:H_AndersonHubbard}. In accordance with our general philosophy, we %want to
formulate this theory completely in real space, so as to keep the advantages discussed in the previous sections.
Let us mention again, however, that this particular choice is motivated by the type of applications we have in mind, and should be reexamined on a case-by-case basis. For example, a formulation in momentum space (the DCA) should be preferred when it is necessary to explicitly maintain translation invariance, and a typical medium formulation should be employed to capture the effects of Anderson localization \cite{Byczuk2005, Braganca2015, Terletska2021, Tam2021}.

Consider first the case with diagonal disorder only. In a disordered system, the MCPA replaces an exact lattice average by the sum over disorder configurations on a finite cluster. As a consequence of interactions, each of these configurations now acquires a self-energy $\underline{\Sigma}^{\mathrm{imp}}$, which we compute by solving the corresponding quantum impurity problem for each disorder configuration. The average can then be determined in the same way as before
\begin{subequations}
\begin{align}\label{eq:C-CDMFT-diagonal1}
    \expval{\underline{G}(\Tilde{\bm{r}}_i, \Tilde{\bm{r}}_i; i\omega_n)} = \expval{\qty[\underline{g}_i^{-1}(i\omega_n) - \underline{\Delta}(i\omega_n) - \underline{\Sigma}^{\mathrm{imp}}(i\omega_n)]^{-1}}.
\end{align}
We now define a new self-energy, $\underline{\Sigma}$, that includes the effects of both interactions and disorder, by writing
\begin{align}
    \expval{\underline{G}(\Tilde{\bm{r}}_i, \Tilde{\bm{r}}_i; i\omega_n)} &= \qty[i \omega_n \, \underline{1} - \underline{\Delta}(i\omega_n) - \underline{\Sigma}(i\omega_n)]^{-1} \label{eq:C-CDMFT-diagonal2} \\
    &= \frac{N_c}{N} \sum_{\Tilde{\bm{k}}} \qty[i\omega_n \, \underline{1} - \underline{t}(\Tilde{\bm{k}}) - \underline{\Sigma}(i\omega_n)]^{-1}. \label{eq:C-CDMFT-diagonal3}
\end{align}
\end{subequations}
Here the last equality expresses the usual self-consistency requirement for the effective medium.
The self-consistent solution of Eqs.~\eqref{eq:C-CDMFT-diagonal1}-\eqref{eq:C-CDMFT-diagonal3} determines the disorder-averaged cluster Green function $\expval{\underline{G}}$ for an interacting system. The method treats both effects on an equal footing, ignoring correlations beyond the size of the chosen cluster, and systematically approaches the exact solution of %to
the problem as $N_c$ is increased, i.e., the algorithm is a controlled approximate technique with the control parameter $N_c$.

Furthermore, the effects of off-diagonal disorder can be straightforwardly incorporated into the method simply by reintroducing the BEB component structure.
Since such a structure is not typically supported by quantum impurity solvers, we here give a simple prescription for treating this extension: for a given realization of disorder, encoded in the locator $\Hat{\underline{g}}_i$, the bare BEB Green function reads
\begin{equation}\label{eq:bare-C-CDMFT}
    \Hat{\underline{G}}^{\mathrm{imp}}_0 = \qty[\Hat{\underline{1}} - \Hat{\underline{g}}_i(i\omega_n) \Hat{\underline{\Delta}}(i\omega_n)]^{-1}\Hat{\underline{g}}_i(i\omega_n).
\end{equation}
Due to the completeness relation $\dyad{A_i} + \dyad{B_i} = \hat{1}$, this quantity reduces to a conventional cluster propagator after summing over its component indices
\begin{equation}\label{eq:bare-CDMFT}
        \underline{G}^{\mathrm{imp}}_0 = \sum_{\mathrm{IJ}} (\Hat{\underline{G}}^{\mathrm{imp}}_0)^{\mathrm{IJ}},
\end{equation}
which can then be used as an input to standard impurity solvers. Finally, the solution of this impurity problem provides the matrix elements $\bra{I_i} G^\mathrm{imp}_{ij} \ket{J_j}$, from which the interacting BEB Green function can then be constructed.

\begin{figure*}[t!]
    \centering
    \includegraphics[width=1\textwidth]{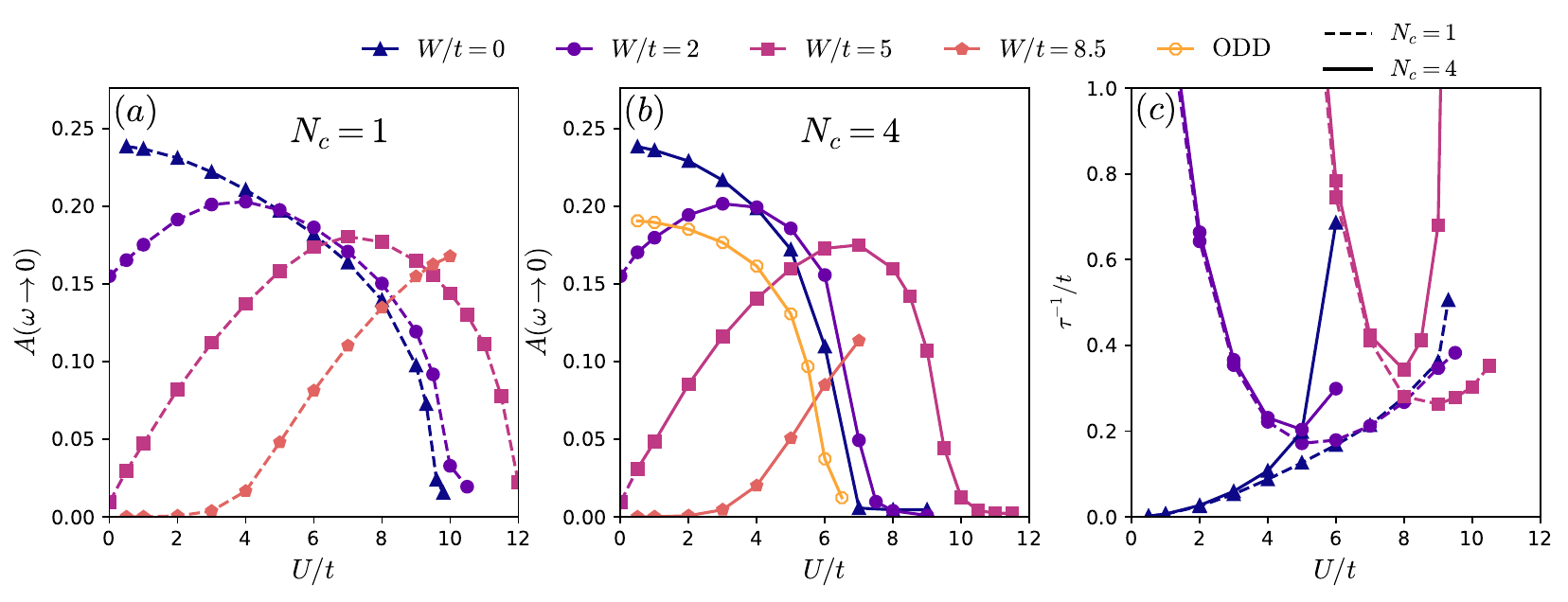}
    \caption{(a) and (b): Matsubara estimate of the local spectral function (density of states) at the Fermi energy as a function of $U/t$ and for various diagonal disorder strengths $W/t$ in single-site (a) (dashed) and four-site C-CDMFT (b) (solid lines). In (b) we also show a calculation with completely off-diagonal disorder (ODD) with $\tau^{\mathrm{AA}} = \tau^{\mathrm{BB}} = 1.37$ and $\tau^\mathrm{AB} = 0.2$, where the Mott transition occurs at a slightly lower $U/t$. In (c) we present the scattering rate (inverse lifetime) $\tau^{-1}$ for the diagonal disorder calculations and both $N_c = 1$ (dashed) and $4$ (solid). In all cases the impurity concentration is $c=0.5$ and the temperature is $T = 0.1 t$.}
    \label{fig:spectral_weights_scattering_rates}
\end{figure*}

We illustrate the full self-consistency flow for this theory, which we term C-CDMFT, in Fig.~\ref{fig:SC_BEB-CDMFT}: starting from an initial guess for the self-energy (e.g., $\hat{\underline{\Sigma}} \equiv 0$), the BEB cluster Green function is computed via Eq.~\eqref{eq:C-CDMFT-diagonal3}. We then use Dyson's Eq.~\eqref{eq:C-CDMFT-diagonal2} to find the corresponding hybridization function $\hat{\underline{\Delta}}$, which in turn fixes the set of bare impurity Green functions $\underline{G}_0^{\mathrm{imp}}$ by means of Eqs.~\eqref{eq:bare-C-CDMFT}-\eqref{eq:bare-CDMFT}. After solving this ensemble of interacting problems (in this work by means of continuous-time quantum Monte Carlo), the resulting Green functions are promoted back to the BEB quantities and configuration averaging is performed. Finally, invoking again Dyson's equation, a new self-energy is extracted and the algorithm repeats until convergence in the quantities $\langle{\hat{\underline{G}}}\rangle, \hat{\underline{\Sigma}}$, and $\hat{\underline{\Delta}}$ is reached.

While interacting models with diagonal and off-diagonal disorder have previously been investigated in infinite dimensions~\cite{Dobrosavljevic1993, Dobrosavljevic1994}, and a single-site version of the above theory has recently been applied to the Bethe lattice~\cite{Weh2021}, our extension of the method provides a systematic route for including important nonlocal correlation effects complementary to the existing DCA framework ~\cite{Ekuma2014a, Ekuma2014b, Terletska2014,Terletska2018,Zhang2016,Tam2021}.
An important advantage of our pure real-space formulation is that spatial modulations resulting, e.g., from antiferromagnetic ordering, are easily included. Furthermore, the solution of each impurity problem corresponds to a different physical lattice environment, which allows us to analyze the importance of impurity arrangements 
to various observables. In Sec. \ref{sec:afm} we apply such a decomposition to the staggered susceptibility, which allows us to directly identify two distinct regimes in the model's response.

We conclude with some practical aspects of the algorithm. While we generally observed a fast convergence of the self-consistency loop Fig.~\ref{fig:SC_BEB-CDMFT} in the regimes investigated here, the need to solve an entire ensemble of interacting impurity problems (their number increases exponentially with the cluster size $N_c$) greatly raises the computational cost of the method. It is therefore crucial to exploit point group symmetries relating disorder configurations to reduce the number of independent computations that have to be performed. For example, when $N_c = 2\times 2 = 4$ only $6$ out of the full $2^4 = 16$ impurity problems fall into different symmetry classes. Especially for larger cluster sizes, we can reduce the cost even further by sampling only a subset of the configurations generated by a Markov chain~\cite{Jarrell2001} (cf. Appendix~\ref{app:Sampling}). Finally, we found it useful to further compress the Green function data generated during the self-consistency cycle using the intermediate representation~\cite{Shinaoka2017, Chikano2019}.
For practical purposes even more important than the proliferation of disorder configurations with cluster size is the numerical cost of solving each individual impurity problem. Disorder-induced correlations beyond computationally feasible sizes of the impurity problem can be captured by subdividing a large unit cell into smaller cluster problems, and then performing the disorder average over the large cell. The number of quantum impurity problems to be solved remains bounded by the possible configurations in the subclusters, while the configurations on the large cluster can be sampled stochastically.

\section{C-CDMFT for the Dirty Mott Transition}
\label{sec:mott}

At low temperatures, the (paramagnetically restricted) DMFT solution of the clean half-filled Hubbard model undergoes a transition from a renormalized metallic state to a correlated (Mott) insulator when the interaction strength $U$ exceeds a critical value $U_{\mathrm{c2}}$~\cite{Georges1996}.
A very different transition takes place in the noninteracting diagonally disordered system with concentration $c=0.5$ equal to the filling: as discussed in Sec.~\ref{sec:cpa}, here a splitting of the density of states into impurity subbands takes place as $W$ exceeds roughly half the bandwidth. In the insulator, sites on which the disorder potential is lower are doubly occupied, and turning on interactions in such a state pushes the occupied states up in energy until the two impurity bands collapse into a single one at $U \sim W$~\cite{Lombardo2006} (see also Appendix \ref{app:HF}).
This metallic-like regime is characterized by strong screening of the disorder potential due to interactions. See, for instance Refs.~\cite{Tanaskovic2003,Poli2025}. A further increase of the repulsion strength then results in a second transition, converting the metal into a Mott insulator.

In order to study this two-step transition in more detail, we show in Fig.~\ref{fig:spectral_weights_scattering_rates} the local spectral weight estimate
\begin{equation}\label{eq:spectral_weight_0}
    A(\omega=0) \approx -\frac{\beta}{\pi} \expval{G_{ii}(\tau = \beta/2)},
\end{equation}
over a range of $U$ and different disorder strengths $W$, both in the single-site approximation $N_c = 1$ as well as for cluster size $N_c = 2 \times 2 = 4$. As before, we fix the concentration at $c = 0.5$ to remain at particle-hole symmetry.

We consider first the limit in which $U$ is small. In accordance with the CPA band-splitting picture, increasing the disorder strength results in a gradual reduction of $A(\omega=0)$, cf.~Fig.~\ref{fig:2d_MCPA_spectra}.
As demonstrated in Sec.~\ref{sec:mcpa}, nonlocal scattering slightly increases the critical disorder strength, but this effect is too small to be observed on the $2\times 2$ cluster.
Turning now to interaction effects, as $U$ increases we indeed observe a strong screening of the impurity potentials, as evidenced by an almost complete restoration of the spectral weight as compared to its clean, noninteracting value. This behavior is consistent with previous studies~\cite{Lombardo2006, Weh2021} and remains largely unaffected by cluster corrections.
However, quantitative differences become apparent in the subsequent Mott transition.
Generically, the need to first overcome the band insulator gap raises $U_\mathrm{c2}$ monotonically as a function of $W$, but as is known from studies of the clean Mott transition, single-site dynamical mean-field theory overestimates the critical value $U_\mathrm{c2}$ of the transition, which can shift to significantly smaller values as nonlocal correlations are taken into account~\cite{Moukouri2001,Schaefer2015,Klett2020,Downey2023,Meixner2024}.
We observe the same effect here, but in a scenario where the transition does not occur out of a simple renormalized metal, but rather a strongly disorder-screened state.
This effect is most clearly seen at $W = 5t$. Here the insulating gap reopens at a repulsion strength of $U \sim 12 t$ in the single-site approximation, while the same transition already takes place at $U \sim 10 t$ on the cluster.
This implies that the range over which the intermediate metallic state can exist becomes substantially narrower as a result of short-ranged dynamical correlations. In both cases the disorder dependence of the spectral weight leads to an interesting counterintuitive effect: a Mott insulator existing close to $U_\mathrm{c2}$, can be turned back into a metallic state by \emph{increasing} the strength of disorder. Consider, 
e.g., the point $U = 7t$ in Fig.~\ref{fig:spectral_weights_scattering_rates} (b) as a function of $W$ (see also \cite{Byczuk2005}).

To investigate the metallic regime in more detail, we study the scattering rate $\tau^{-1}_{k_F}$, extracted from the averaged Green function via \footnote{On the cluster the scattering rate picks up a (weak) momentum-dependence. Since our real-space embedding partially breaks translation invariance, we first reperiodize the self-energy. For details, see~\cite{Sakai2012}.}:
\begin{equation}
  \frac{1}{2\tau_{\bm{k}_F}} = Z_{\bm{k}} \Im \Sigma_{\bm{k}_F}(i0^+),
\end{equation}
with
\begin{equation}
    Z_{\bm{k}}^{-1} = 1 - \eval{\frac{\partial \Im \Sigma_{\bm{k}_F}}{ \partial \omega_n}}_{\omega_n\rightarrow i0^+}.
\end{equation}
A new feature, absent in single-site calculations, is that this quantity possesses a momentum dependence on the cluster. This may, for example, induce a pseudogap structure in the low-frequency density of states \cite{Parcollet2004}.
In our calculations, however, we do not observe any appreciable $\bm{k}$-dependence in $\tau^{-1}_{\bm{k}}$ along the Fermi surface, implying that the cluster corrections we observe remain essentially quantitative at this temperature.

We show the scattering rate averaged over the Fermi surface in Fig.~\ref{fig:spectral_weights_scattering_rates} (c), for both single-site and cluster calculations. Consistent with the behavior of the spectral weight, $\tau^{-1}_{k_F}$ in disordered systems is very large at weak interaction, 
but then reduces strongly and reaches a minimum for $U \sim W$ as a consequence of screening, and in agreement with the zero temperature single-site results of Ref.~\cite{Tanaskovic2003}. The minimum is rather broad for $N_c = 1$, and significantly narrower for $N_c = 4$.
Also here, the nonmonotonic dependence of the scattering rate creates a scenario where an increase in disorder strength (close to the Mott transition) makes the system apparently more coherent. Note, however, that we cannot directly address localization effects within our method. While the scattering rate in the intermediate state is small, the conductivity might still be zero.

\begin{figure*}[t!]
    \centering
    \includegraphics[width=\textwidth]{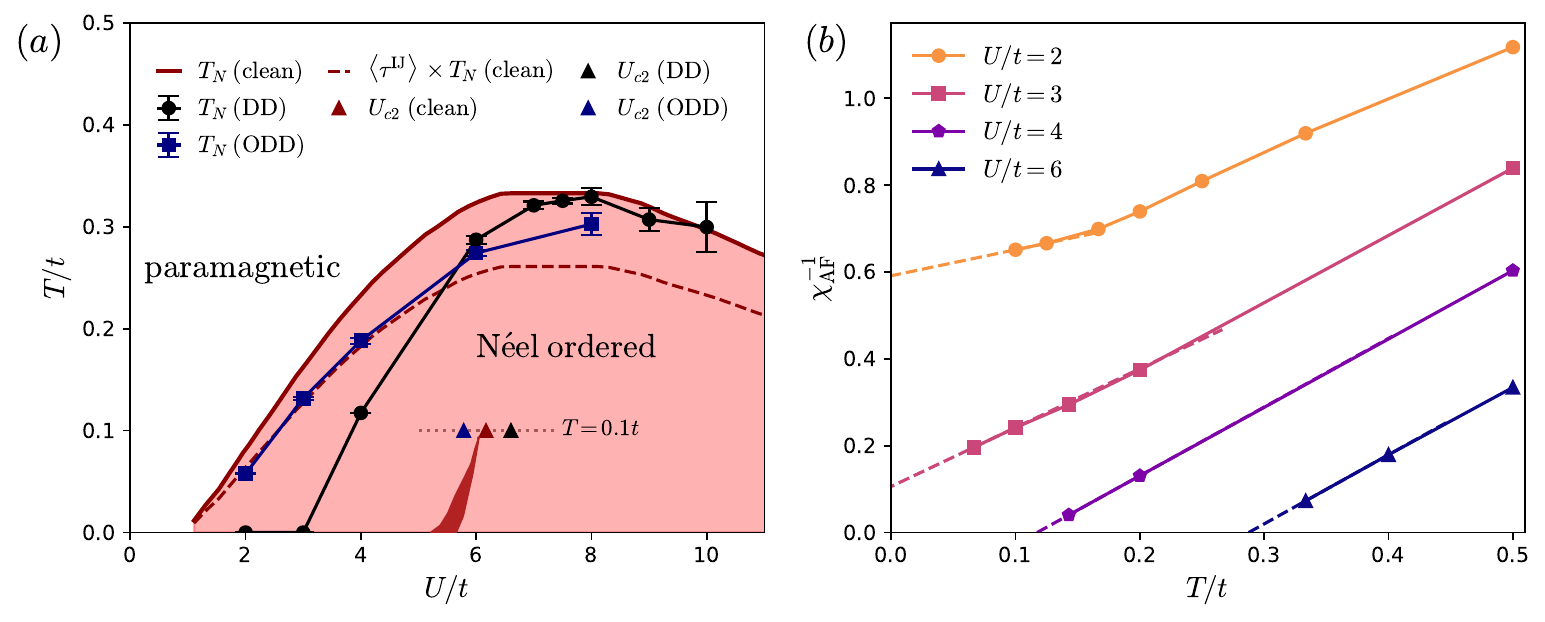}
    \caption{(a) Magnetic C-CDMFT phase diagram on the simple square lattice (with cluster size $N_c = 2\times 2$), showing the Néel temperatures $T_N$ for the clean system (reproduced from~\cite{Klett2020}), diagonal disorder (DD) $W/t=2$, and off-diagonal disorder (ODD) $\tau^\mathrm{AA} = \tau^\mathrm{BB} = 1.37$, $\tau^\mathrm{AB} = 0.2$. At weak coupling, the ODD ordering temperature is identical to that of the clean system rescaled by the averaged hopping amplitude $\expval{\tau^\mathrm{IJ}}$ (dashed line). Error bars are estimated by averaging over the last five iterations of the self-consistency cycle. At temperature $T=0.1t$ we also show the critical couplings $U_\mathrm{c2}/t$ for the paramagnetic Mott transition for the different types of disorder, extracted from Fig.~\ref{fig:spectral_weights_scattering_rates} (b) (triangles). The full temperature dependence of the transition region (dark red shaded area) is reproduced from~\cite{Park2008}
    (b) Inverse susceptibilities $\chi^{-1}_\mathrm{AF} = \qty(\dd{m_\mathrm{AF}}/\dd{B})^{-1}$ for diagonal disorder. The Néel temperatures in (a) are obtained by a linear extrapolation of $\chi^{-1}_\mathrm{AF}$ (dashed lines). For $U/t = 2$ and $U/t = 3$ the susceptibility does not diverge at any nonzero temperature, hence the system remains paramagnetic.}
	\label{fig:2d_magnetic_phase_diagram}
\end{figure*}

Lastly, we also performed a representative cluster calculation for purely off-diagonal disorder $\tau^\mathrm{AA} = \tau^{\mathrm{BB}} = 1.37$, $\tau^{\mathrm{AB}} = 0.2$, shown in Fig.~\ref{fig:spectral_weights_scattering_rates} (b). The hopping parameters were chosen such as to leave the noninteracting bandwidth approximately unchanged. No band-splitting transition occurs in this case, so no increase of $U_{\mathrm{c2}}$ should be expected. Interestingly, we instead note a small reduction of the metallic regime, i.e., the Mott state is apparently stabilized by the presence of different hopping scales. We will discuss this point further in the following section.

\section{C-CDMFT and Disorder Diagnostics for Antiferromagnetism}
\label{sec:afm}
On the simple square lattice, and at sufficiently low temperatures, the spatial mean-field nature of any DMFT cluster theory with finite $N_c$ results in antiferromagnetic ordering, even in 2 dimensions. The associated Néel temperature $T_N$ sets the scale at which antiferromagnetic correlations become large, i.e., comparable to the cluster size. To study how this scale is affected by adding disorder, we compute the staggered susceptibility $\chi_{\mathrm{AF}}$ in the paramagnetic parent state as a function of temperature. This quantity is obtained by applying a small staggered field $B$ and measuring the magnetic response $\chi_{\mathrm{AF}} = \dd{m_\mathrm{AF}}/\dd{B}$. In the vicinity of the transition, the susceptibility diverges according to mean-field theory $\chi_{\mathrm{AF}}\propto \left|T-T_N\right|^{-1}$. We thus determine $T_\mathrm{N}$ by extrapolating to the singularity.

Figure~\ref{fig:2d_magnetic_phase_diagram} presents our results for the magnetic phase diagram, computed within $2\times 2$ C-CDMFT, both for completely diagonal ($W=2t$) and purely off-diagonal disorder (for the same parameters as in Fig.~\ref{fig:spectral_weights_scattering_rates} (b)). For reference we also show the phase boundary in the clean limit, reproduced from~\cite{Klett2020}: due to nesting at half-filling, the magnetic dome extends over the entire range of $U$ values (only $U \geq 1t$ is shown here), but the underlying mechanisms at weak and strong coupling are quite different~\cite{Fratino2017}.
An important reference scale for this crossover is the location of the paramagnetic Mott metal-insulator transition, obtained from the metastable paramagnetic solution of the C-CDMFT calculation (Sec.~\ref{sec:mott}) \footnote{We determine $U_\mathrm{c2}$ from the inflection point of the spectral weight curves, $[A(\omega=0)](U)$}.
When $U \gtrsim U_\mathrm{c2}$, it becomes appropriate to describe the physics in terms of localized --- instead of itinerant --- magnetic moments, and we expect these two regimes to respond very differently to the different types of disorder. While magnetic phase boundaries can also be computed in a single-site method, our having access to extended real space disorder configurations enables us to investigate this response explicitly. In the following we will discuss the cases of strictly diagonal and off-diagonal disorder in more detail.

\begin{figure*}[t!]
    \centering
    \includegraphics[width=1\textwidth]{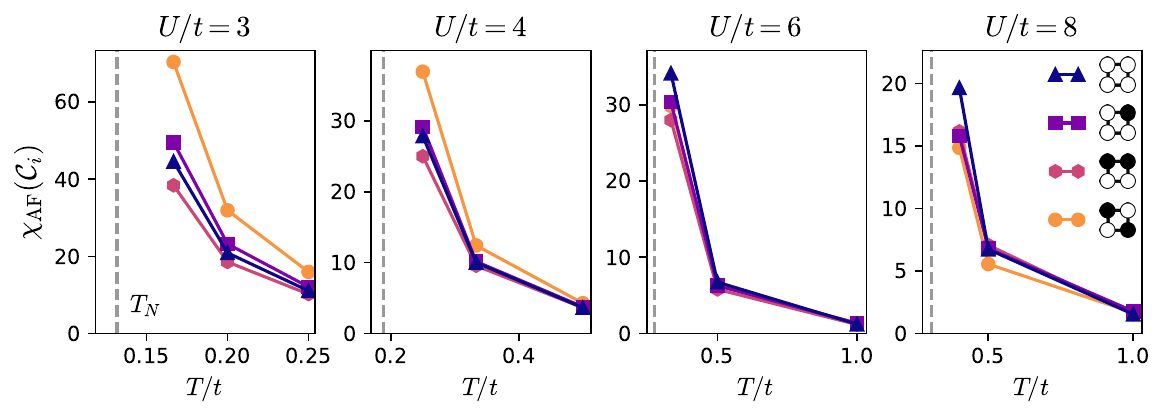}
    \caption{Disorder diagnostics for off-diagonal disorder: contributions $\chi_\mathrm{AF}(\mathcal{C}_i)$ from different disorder configurations to the susceptibility $\chi_\mathrm{AF}$. The Néel temperature $T_N$ (dashed line) corresponding to the phase boundary in Fig.~\ref{fig:2d_magnetic_phase_diagram} (a) is also shown. At weak coupling $U/t \lessapprox 5$, the configuration with strongest response features only \emph{weak bonds} $\tau^\mathrm{AB}=0.2$. Conversely, in the strong coupling regime $U/t \gtrapprox 5$ it is the configuration with only \emph{strong bonds} $\tau^\mathrm{AA} = \tau^\mathrm{BB} = 1.37$ that represents the largest contribution to $\chi_\mathrm{AF}$.}
    \label{fig:config_susceptibilities}
\end{figure*}

\subsection{Diagonal disorder}

At weak coupling the magnetic state may be understood within a simple band picture with a sublattice structure and a gap $\sim m_\mathrm{AF} U$.
Including diagonal disorder then has the same effect as in the nonmagnetic system, namely a splitting of each band (see Appendix \ref{app:HF}).
This reduces the gap by an amount $\sim W$ and finally destroys the antiferromagnet once $W \gtrsim U$.
In accordance with this picture, our numerical results for the magnetic dome at disorder strength $W = 2t$ show a suppression of the magnetic order up to a quantum critical point at $U \sim W$, indicating frustration due to the inhomogeneities.
In this regime, the staggered susceptibility deviates from the expected Curie law at low temperatures, preventing divergence.
The linear dependence $\chi^{-1} \propto T - T_\mathrm{N}$ is restored only for stronger interactions, and antiferromagnetic order sets in.
The corresponding Néel temperature first increases as a function of $U$, until it reaches a maximum in the vicinity of $U_\mathrm{c2}$.
Interestingly, we do not observe any appreciable reduction of the maximal value of the transition temperature as compared to the clean limit --- again consistent with strong screening of the disorder potential in the vicinity of $U_\mathrm{c2}$.
Rather, the two magnetic domes become almost identical for $U > U_\mathrm{c2}$, and decrease with the common exchange energy $J = t^2/U$ of localized magnetic moments.
Hence the most striking effects of diagonal disorder on the magnetic phase occur at weak coupling, where the suppression of magnetic correlations results in a tunable quantum critical point at $U \sim W$.
These results agree well with previous studies of an infinite dimensional model~\cite{Janis_1993, Ulmke1995} and direct quantum Monte Carlo simulations~\cite{Ulmke1997}.

\subsection{Off-diagonal disorder and disorder diagnostics}
The situation is qualitatively different when the disorder is off-diagonal. Considering first the large-$U$ regime in which the magnetic order is due to the physics of localized moments, randomness in the hopping amplitudes also implies a distribution of superexchange couplings, whose average $\expval{J} \sim \big\langle \qty|\tau^{\mathrm{IJ}}|^2 \big\rangle {4t^2}/{U}$ is set by the second moment of $\tau^{\mathrm{IJ}}$. On the other hand, the presence of weaker bonds in the system can effectively raise the local correlation scale $U/t$, favoring the formation of local moments~\cite{Milovanovic1989}. This latter effect explains the decrease of $U_{\mathrm{c2}}$ as compared to the clean Mott transition we observe in Fig.~\ref{fig:spectral_weights_scattering_rates}, since the transition is driven by the formation of local moments.

Our results for the magnetic dome with off-diagonal randomness are shown in Fig.~\ref{fig:2d_magnetic_phase_diagram}(a). Interestingly, even though disorder is expected to spoil the Fermi surface nesting and van Hove singularity responsible for the weak-coupling instability in the clean system, we observe the onset of magnetic order down to the lowest interaction strength investigated. Moreover, up to the value $U_{\mathrm{c2}}$ obtained for this model in the previous section, the magnetic phase boundary is remarkably well described by a simple rescaling of the clean magnetic dome by the \emph{average} hopping amplitude $\expval{\tau^\mathrm{IJ}}$. This is consistent with previous numerical evidence~\cite{Ulmke1997}, suggesting that off-diagonal disorder alone is insufficient for lifting the weak-coupling instability. Hence we believe that the quantum critical point in this scenario remains fixed at $U=0$.

Deviations from this simple scaling appear for $U > U_{\mathrm{c2}}$. As stated above, the relevant scale in this regime is the second moment $\big\langle \qty|\tau^{\mathrm{IJ}}|^2 \big\rangle$ (which is $\approx 1$ for our choice of parameters) instead of the average. To further investigate this crossover, we decompose the staggered susceptibility $\chi_{\mathrm{AF}}$ into contributions from the distinct impurity configurations on the cluster. Performing such a \emph{disorder diagnostics} is very natural within our real-space framework, where the susceptibility is explicitly computed according to
\begin{equation}\label{eq:chi_sum}
    \chi_{\mathrm{AF}} = p(\mathcal{C}_1) \chi_{\mathrm{AF}}(\mathcal{C}_1) + p(\mathcal{C}_2) \chi_{\mathrm{AF}}(\mathcal{C}_2) + \dots,
\end{equation}
which is a sum over the susceptibilities $\chi_{\mathrm{AF}}(\mathcal{C}_i)$ of all inequivalent configurations $\{ \mathcal{C}_i \}$ embedded in the self-consistently determined medium, and $p(\mathcal{C}_i)$ is the probability for these configurations to occur in the average. The quantities $\chi_\mathrm{AF}(\mathcal{C}_i)$ are shown in Fig.~\ref{fig:config_susceptibilities} (see Appendix~\ref{app:diagnostics} for additional plots). There are four distinct configurations on the $N_c = 2 \times 2$ cluster, and we will focus on the two most important ones: (i) all sites are of the same type and hence connected through only strong bonds $\tau^{\mathrm{AA}} = \tau^{\mathrm{BB}} = 1.37$, and (ii) a checkerboard arrangement of A and B sites, resulting in only weak bonds $\tau^{\mathrm{AB}} = 0.2$. In the weak-coupling limit, the latter configuration is expected to respond more strongly to applied magnetic fields, as the interaction strength $U/t$ in such an arrangement is effectively raised. On the other hand, the presence of strong bonds increases the superexchange scale $t^2/U$, and is therefore more susceptible in the strong-coupling limit.
In excellent agreement with this picture, we find that the largest local response $\chi_\mathrm{AF}(\mathcal{C}_i)$ in
Eq.~\eqref{eq:chi_sum} corresponds to configuration (i) for $U \lesssim U_\mathrm{c2}$, but then switches to configuration (ii) for $U \gtrsim U_\mathrm{c2}$.

Thus, C-CDMFT provides remarkable insights into the physical properties of the underlying phase that go well beyond what can be extracted from single-site theories. The disorder configurations sampled during the self-consistency cycle act as local perturbations relative to the averaged environment and therefore contain information about the system's response. Furthermore, the resolution of this diagnostics tool is naturally refined by increasing the cluster size, thereby giving access to a larger number of impurity arrangements and longer-ranged spatial correlations. Moreover, this analysis is not restricted to magnetic susceptibilities and can be straightforwardly extended to probe, e.g., superconducting correlations. This is especially interesting in connection with experiments on cuprate and nickelate superconductors, in which multiple competing instabilities have been reported. This analysis will be left for future work.

Finally, returning to the decomposition Eq.~\eqref{eq:chi_sum}, we also note that the configuration with two strong and two weak bonds, favoring the formation of nonmagnetic spin-singlets, shows a comparatively weak response over the entire range of interactions.
In fact, once the variance in the exchange couplings $\expval{J^2} - \expval{J}^2$ exceeds a critical value, such singlets have been shown to completely destroy the antiferromagnetic ordering in the local moment regime~\cite{Bhatt1982, Kirckpatrick1996, Ulmke1997, Zhou2009}.
While our choice of parameters does not allow us to investigate such effects in the present work, it is nonetheless interesting to observe that this behavior is exactly opposite to that of diagonal disorder, where the frustration mechanism is most effective in the weak-coupling regime.

\section{Conclusions}
\label{sec:conclusions}

We constructed an embedded quantum cluster theory, C-CDMFT, by combining the molecular coherent potential approximation in its BEB formulation with the cellular dynamical mean-field theory. The approach, which is formulated entirely in real-space, treats exactly both disorder and dynamical fluctuations on short length scales and on an equal footing, complementing currently available techniques. At the cost of explicitly breaking translation invariance, it has important advantages over the DCA approach when it comes to studying the system's \emph{local} properties: on the one hand, (translation)symmetry-breaking fields are straightforwardly incorporated, as exemplified by our study of the antiferromagnetic instability, while on the other, and more importantly, the ensemble of quantum impurity models that is solved in the process \emph{is physical}. This allows us to decompose and diagnose averages into contributions from various disorder arrangements, which play the role of intrinsic local perturbations, and provide deep insights into the correlation structure of the underlying state.

In this work we presented first results obtained with our method, both on the influence of disorder on the Mott transition, and the stability of the magnetic phase. Our results agree with and extend important early and recent work, paving the way to exploring previously inaccessible nonlocal effects.
Moreover, the possibility 
to include diagonal as well as off-diagonal disorder makes the method applicable to sufficiently realistic descriptions of many strongly interacting, chemically doped compounds. In this context, the disorder diagnostics promises an interesting route to the study of competing instabilities, particularly in relation to superconductivity, which bears directly on recent experimental questions.

\begin{figure}[t!]
    \centering
    \includegraphics[width=0.99\columnwidth]{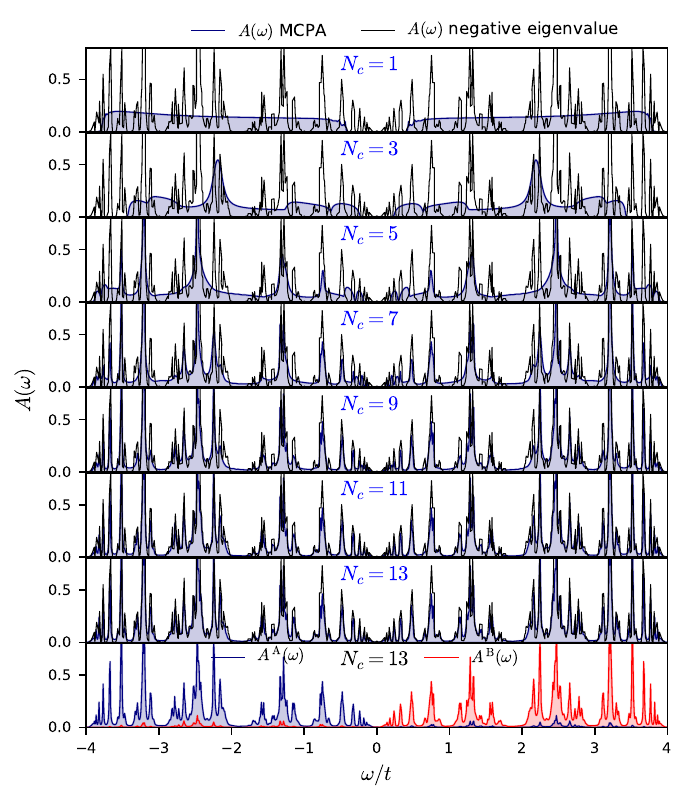}
    \caption{Local spectral weight (density of states) $A(\omega)$ of a disordered $1d$ chain. In blue: MCPA results for cluster sizes $N_c \in \{1, 3, 5, 7, 9, 11, 13\}$. Black: numerically exact data based on the negative eigenvalue theorem \cite{Dean1972} with energy resolution $\Delta \omega /t  = 0.01$ (reproduced from~\cite{Rowlands2006}). The impurity parameters are $W/t = 4$, $c=0.5$. Last row shows the corresponding BEB decompositions.}
    \label{fig:1d_MCPA_spectra}
\end{figure}
\begin{acknowledgements}
\textit{Acknowledgements.}
We thank Vladimir Dobrosavljevi{\'c}, Kilian Fraboulet, Karsten Held, Thomas Maier, Michael Meixner, M{\'a}rio Malcolms de Oliveira, Hanna Terletska, and Roser Valent{\'i} for useful discussions. The authors acknowledge the computer support teams at CPHT {\'E}cole Polytechnique and the computer service facility of the MPI-FKF for their help. P.T. acknowledges support from the CCQ-Columbia University Joaquin Luttinger fellowship.
S.A. acknowledges financial support from the Deutsche Forschungsgemeinschaft (DFG)
within the research unit
FOR 5413/1 (Grant No. 465199066).
This research was also funded 
by the Austrian Science Fund (FWF) 10.55776/I6946.
\end{acknowledgements}

\appendix
\section{Computational effort}
\label{app:comp_cost}
To give an estimate of the computational effort, a C-CDMFT sample calculation on the $N_c=2\times 2$ cluster for the half-filled square lattice with both diagonal and off-diagonal disorder at $U/t=1$ and $T=0.05t$ requires roughly 15 iterations with 300 core-hours each until satisfactory convergence of the averaged quantities is reached. This implies a cost of approximately 5000 core-hours for a single data point.

\begin{figure}[t!]
    \centering
    \includegraphics[width=0.5\textwidth]{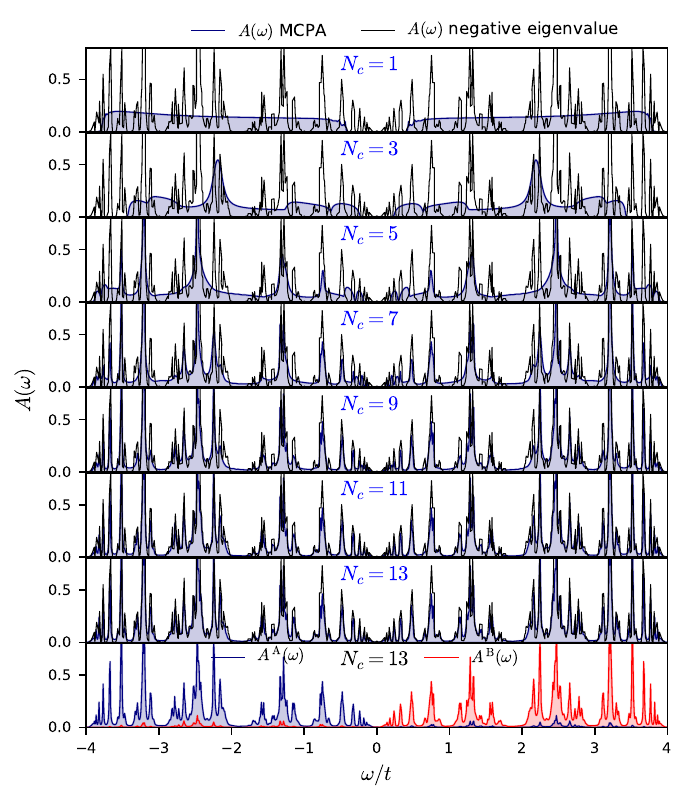}
    \caption{Local spectral weight (density of states) $A(\omega)$ of a disordered $1d$ chain. In blue: MCPA results for cluster sizes $N_c \in \{1, 3, 5, 7, 9, 11, 13\}$. Black: numerically exact data based on the negative eigenvalue theorem \cite{Dean1972} with energy resolution $\Delta \omega /t  = 0.01$ (reproduced from~\cite{Rowlands2006}). The impurity parameters are $W/t = 4$, $c=0.5$. Last row shows the corresponding BEB decompositions.}
    \label{fig:1d_MCPA_spectra}
\end{figure}

\section{Supplemental plots}
\label{app:supplemental_plots}

In this section we present additional figures and calculations. 

\subsection{Results in one dimension}
\label{app:1dresults}

To underline the convergence properties of the MCPA algorithm (i.e., C-CDMFT at $U=0$), we performed calculations on the one-dimensional Anderson model with diagonal disorder strength $W = 4t$ and concentration $c=0.5$ on clusters up to $N_c = 13$ sites (Fig.~\ref{fig:1d_MCPA_spectra}). Exact spectral functions for this parameter set have previously been reported in the literature~\cite{Rowlands2006}, which we reproduce here to draw comparisons with our results. We first note that the CPA estimate ($N_c=1$) incorrectly predicts a gap around $\omega = 0$, and misses essentially all features in the spectrum. The latter are associated with coherent scatterings between different sites. As such, increasing the cluster size systematically introduces peaks at positions corresponding to the sharp resonances found in the numerically exact solution. At the same time, the gap is successively filled in. For $N_c \sim 10$ we already reach excellent agreement with the exact data: the positions of all peaks are correctly reproduced, and differences in the peak heights are mainly due to numerical artifacts. In addition, we can also exploit the BEB structure built into our algorithm to decompose the spectrum into contributions from the individual components (A and B sites). This is shown in the bottom panel of Fig.~\ref{fig:1d_MCPA_spectra}. As expected, the two components are almost completely separated by the large disorder potential, but spectral weight of the A sites can still be found deep within the spectrum of B states and vice versa. 

\subsection{Hartree-Fock calculations}
\label{app:HF}

\begin{figure}
    \centering
    \includegraphics[width=.5\textwidth]{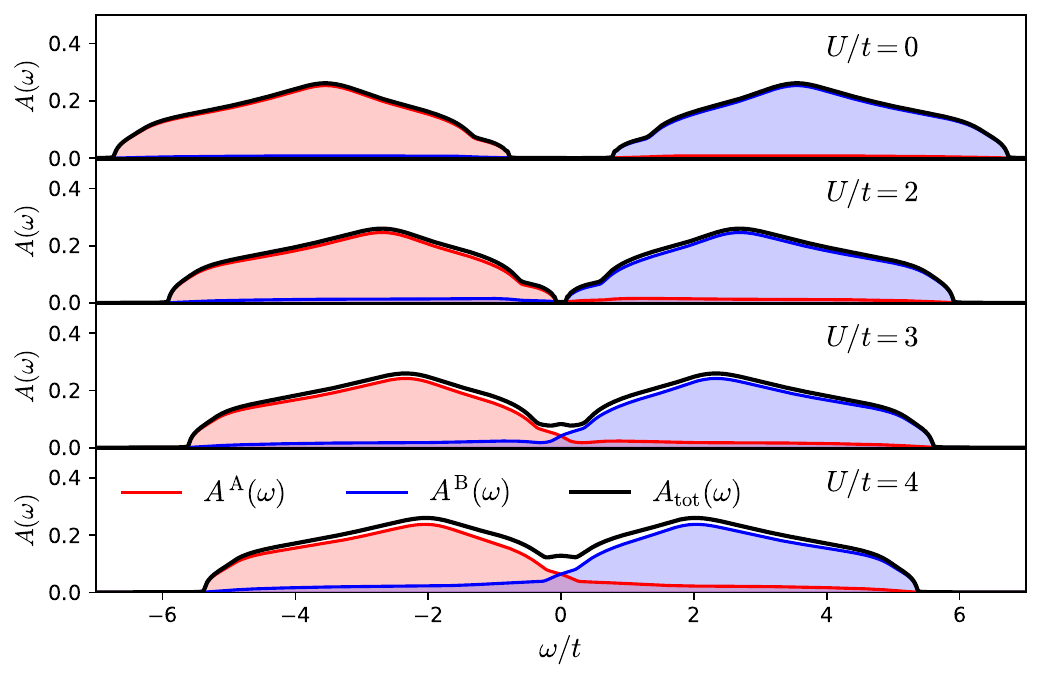}
    \caption{Component resolved C-CDMFT spectral functions $A^\mathrm{A}(\omega)$, $A^\mathrm{B}(\omega)$ and their sum $A_\mathrm{tot} = A^\mathrm{A} + A^\mathrm{B}$ for a $2\times 2$ cluster with diagonal disorder $W/t=7$ and concentration $c=0.5$. The Hartree approximation is used to treat the Hubbard interaction $U$ at temperature $T = 0.1 t$.}
    \label{fig:Hartree_Bandinsulator}
\end{figure}

As pointed out in Secs.~\ref{sec:mott} and~\ref{sec:afm}, the strong disorder - weak interaction limit $W \gg U$ can be understood qualitatively in a band theory picture (see also~\cite{Weh2021} for a similar discussion of the Hartree shift). We therefore show in Fig.~\ref{fig:Hartree_Bandinsulator} C-CDMFT spectral functions where interactions are treated approximately within the static Hartree-Fock approximation. The diagonal disorder strength is fixed at $W = 7t$, strong enough to split the bands at $U=0$, and we show the component-resolved spectral functions for a range of $U$ values. Since the lower band corresponds to doubly occupied sites, the Hartree shift increases their energy and thereby reduces the band gap, until the bands start overlapping at $U \sim W$. While the Hartree treatment becomes unreliable in this regime, this simple picture reproduces well the initial increase in $A(\omega=0)$ shown in Fig.~\ref{fig:spectral_weights_scattering_rates} (a) and (b).

\begin{figure}
    \centering
    \includegraphics[width=.5\textwidth]{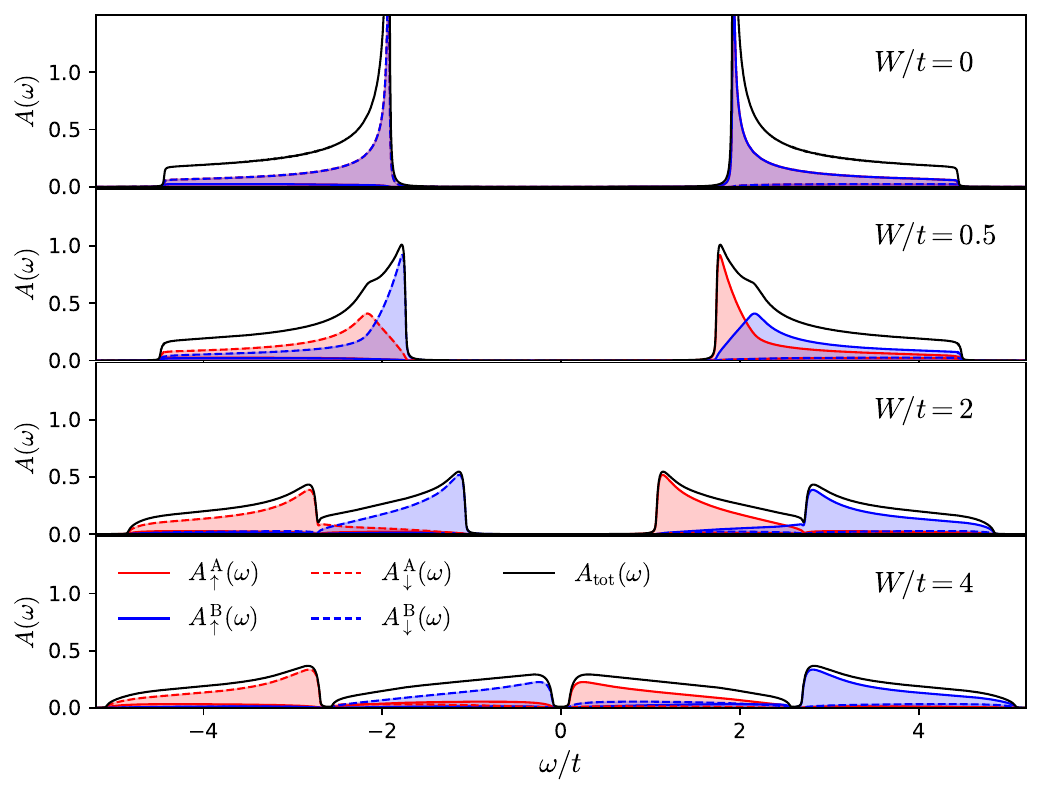}
    \caption{Component and spin resolved local spectral functions $A^\mathrm{I}_\sigma(\omega)$ on a $2\times 2$ cluster displaying antiferromagnetic order. The interaction $U/t = 5$ is treated within the Hartree approximation and various values of the diagonal disorder strength $W/t \in \{ 0, 0.5, 2, 4\}$ are considered. Calculations are performed at a small temperature $T = 0.1t$.}
    \label{fig:Hartree_AF}
\end{figure}

Similarly, to understand the breakdown of antiferromagnetic ordering at weak coupling, we present in Fig.~\ref{fig:Hartree_AF} results from a spin-symmetry broken Hartree-Fock calculation. Here we fix $U = 5t$ (to make the gap more easily visible, $U$ should still be thought of as being small) and consider a sequence of increasing disorder strengths $W < U$. For $W=0$ a mean-field gap $U m_\mathrm{AF}$ with diverging density of states at the inner band edges is formed. Introducing disorder removes this singularity and starts splitting each magnetic subband into two, corresponding to a separation of the onsite energies on the two magnetic sublattices. Again, as $W \sim U$ the gap closes and the antiferromagnetic solution becomes unstable, consistent with the full C-CDMFT result shown in Fig.~\ref{fig:2d_magnetic_phase_diagram}.

\subsection{Disorder diagnostics}
\label{app:diagnostics}

\begin{figure}[b]
    \centering
    \includegraphics[width=.5\textwidth]{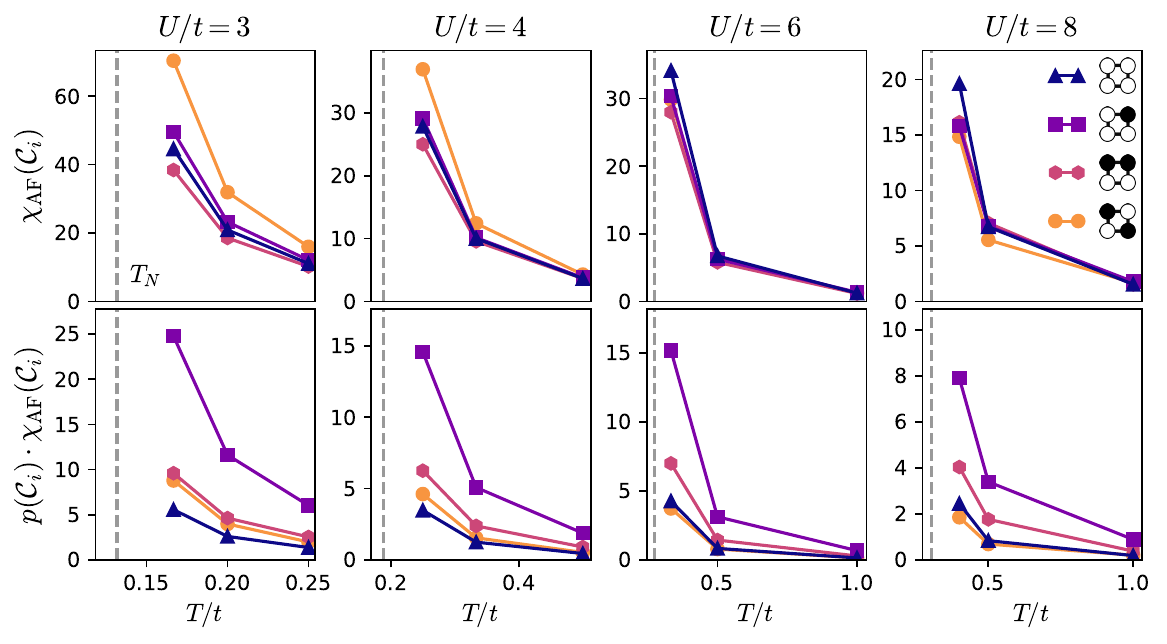}
    \caption{Comparison of the local contributions $\chi_\mathrm{AF}(\mathcal{C}_i)$ (upper row) and their weighted contributions $p(\mathcal{C}_i) \chi_\mathrm{AF}(\mathcal{C}_i)$ to the average Eq.~\eqref{eq:chi_sum} (bottom row). While the local response reveals a change in behavior, the average remains dominated by the \emph{most likely} arrangement of impurities.}
    \label{fig:configs2}
\end{figure}

In Sec.~\ref{sec:ccdmft} we presented the susceptibilities $\chi_\mathrm{AF}(\mathcal{C}_i)$ for individual disorder configurations $\mathcal{C}_i$ and showed that their response to external fields changes qualitatively as interactions drive the system from the weak to strong-coupling regime. In Fig.~\ref{fig:configs2} we show in addition their weighted contribution to the average susceptibility, i.e., $p(\mathcal{C}_i)\chi_\mathrm{AF}(\mathcal{C}_i)$. For concentration $c=0.5$, arrangements are symmetric under the exchange of host and impurity sites, so configurations (i) and (ii) (only strong/weak bonds, respectively) occur twice, while the other configuration with two impurities appears four times, and the single-impurity configuration eight times. The latter thus dominates the averaged response over the entire range of interaction strengths investigated, demonstrating that the change in the local response does not significantly affect its average.

\section{Markov chain configuration sampling}
\label{app:Sampling}

The number of disorder configurations on a cluster grows exponentially with cluster size $N_c$. Particularly in dimensions $d>1$, it therefore becomes necessary to find an accurate approximation to the full average Eq.~\eqref{eq:MCPA1} involving only a small number $\mathcal{N} \ll 2^{N_c}$ of configurations:
\begin{equation}\label{eq:MC_approx}
    \expval{\underline{G}(\omega)} = \sum\limits_\mathcal{C} p(\mathcal{C})\, \underline{G}(\omega ; \mathcal{C}) \approx \frac{1}{\mathcal{N}} \sum\limits_{i=1}^\mathcal{N} \underline{G}(\omega ; \mathcal{C}_i).
\end{equation}
Here the first sum runs over all impurity configurations $\mathcal{C}$, $p(\mathcal{C})$ is the probability of such a configuration and $\underline{G}(\omega; \mathcal{C})$ denotes the corresponding cluster Green function.

A simple protocol for choosing the most relevant configurations is the Metropolis-Hastings algorithm~\cite{Metropolis, Hastings}. Here we give an elementary description of this procedure:
\begin{itemize}
    \item Initialize the Markov chain with an arbitrary configuration $\mathcal{C}_k$, corresponding to a specific realization of $M\leq N_c$ impurities on the cluster. For a binary alloy, the probability of such a configuration is $p(\mathcal{C}_i) = c^M \, (1-c)^{N_c - M}$.
    \item The Markov step is split into two parts. First, a new configuration $\mathcal{C}^\mathrm{prop}_{k+1}$ is proposed with probability  $W^\mathrm{prop}(\mathcal{C}_k \rightarrow \mathcal{C}_{k+1}^\mathrm{prop})$.  A second probability $W^\mathrm{acc}(\mathcal{C}_k \rightarrow \mathcal{C}_{k+1}^\mathrm{prop})$ then determines whether the move $\mathcal{C}_k \rightarrow \mathcal{C}_{k+1}^\mathrm{prop}$ will be accepted or rejected. Here we choose new configurations by a simple flip operation, converting a host site into an impurity and vice versa. The site to be flipped is chosen with uniform probability $W^\mathrm{prop}(\mathcal{C}_k \rightarrow \mathcal{C}_{k+1}^\mathrm{prop}) = 1/N_c$, so that
    \begin{equation}\label{eq:Metropolis_Hastings}
        W^\mathrm{acc}(\mathcal{C}_k \rightarrow \mathcal{C}_{k+1}^\mathrm{prop}) = \min\qty(1, \frac{p(\mathcal{C}_{k+1}^\mathrm{prop})}{p(\mathcal{C}_k)}),
    \end{equation}
    depends only on the ratio $p(\mathcal{C}_{k+1}^\mathrm{prop}) / p(\mathcal{C}_{k})$.
    The next configuration in the Markov chain is $\mathcal{C}_{k+1} = \mathcal{C}^\mathrm{prop}_{k+1}$ with probability \linebreak $W^\mathrm{prop}(\mathcal{C}_k \rightarrow \mathcal{C}_{k+1}^\mathrm{prop}) \cdot W^\mathrm{acc}(\mathcal{C}_k \rightarrow \mathcal{C}_{k+1}^\mathrm{prop})$, and it remains $\mathcal{C}_{k+1} = \mathcal{C}_{k}$ otherwise.

\end{itemize}

The sampling scheme can be further optimized by exploiting exact symmetry properties of the disorder-averaged cluster Green function $\expval{\underline{G}(\omega)}$. Consider, for instance, a $2\times 2$ cluster containing a single impurity. There are four such configurations, each breaking symmetries of the lattice (e.g., invariance under $90^\circ$ rotation). Nevertheless, the average in Eq.~\eqref{eq:MC_approx} is fully symmetric as each of these configurations enters with equal weight. Now, in order for the Monte Carlo approximation to reproduce this property, it is clear that the quantities being sampled in the Markov chain should not be individual disorder configurations, but should rather be grouped into equivalence classes of configurations related by operations of the lattice point group. The advantages of such a sampling procedure are twofold: first, it guarantees that $\expval{\underline{G}(\omega)}$ obeys exacts symmetries of the lattice, and second, it reduces the number of Green functions $\underline{G}(\omega ; \mathcal{C}_i)$ that need to be computed at fixed number $\mathcal{N}$ of configurations. This greatly reduces the computational cost in the case of interacting disordered systems, where computing Green functions is numerically very demanding.

\bibliography{bibliography.bib}

\end{document}